\documentclass[sigconf]{acmart}

\newif\ifanonymous
\anonymousfalse

\AtBeginDocument{%
  \providecommand\BibTeX{{%
    \normalfont B\kern-0.5em{\scshape i\kern-0.25em b}\kern-0.8em\TeX}}}

\setcopyright{acmcopyright}
\copyrightyear{2021}
\acmYear{2021}
\acmDOI{10.1145/1122445.1122456}

\acmConference[Woodstock '18]{Woodstock '18: ACM Symposium on Neural
  Gaze Detection}{June 03--05, 2018}{Woodstock, NY}
\acmBooktitle{Woodstock '18: ACM Symposium on Neural Gaze Detection,
  June 03--05, 2018, Woodstock, NY}
\acmPrice{15.00}
\acmISBN{978-1-4503-XXXX-X/18/06}

\setcopyright{none}
\renewcommand\footnotetextcopyrightpermission[1]{}
\usepackage{listings}
\usepackage{acronym}
\usepackage{csvsimple}
\usepackage{adjustbox}
\usepackage{xcolor}
\usepackage{float}
\usepackage{makecell}
\usepackage{subfig}
\usepackage{tikz}
\usepackage{amsmath}
\usepackage{comment}
\usepackage{multirow}
\usepackage[colorinlistoftodos, textwidth=2.0cm]{todonotes}
\graphicspath{ {./images/} }

\usepackage{url}
\usepackage{breakurl}

\newcommand{\sys}{ESCORT} 
\newcommand{\datatool}{ContractScraper}

\colorlet{punct}{red!60!black}
\definecolor{background}{HTML}{EEEEEE}
\definecolor{delim}{RGB}{20,105,176}
\colorlet{numb}{magenta!60!black}
\lstdefinelanguage{json}{
    basicstyle=\normalfont\ttfamily,
    numbers=left,
    numberstyle=\scriptsize,
    stepnumber=1,
    numbersep=7pt,
    showstringspaces=false,
    breaklines=true,
    frame=lines,
    literate= 
     *{0}{{{\color{numb}0}}}{1}
      {1}{{{\color{numb}1}}}{1}
      {2}{{{\color{numb}2}}}{1}
      {3}{{{\color{numb}3}}}{1}
      {4}{{{\color{numb}4}}}{1}
      {5}{{{\color{numb}5}}}{1}
      {6}{{{\color{numb}6}}}{1}
      {7}{{{\color{numb}7}}}{1}
      {8}{{{\color{numb}8}}}{1}
      {9}{{{\color{numb}9}}}{1}
      {:}{{{\color{punct}{:}}}}{1}
      {,}{{{\color{punct}{,}}}}{1}
      {\{}{{{\color{delim}{\{}}}}{1}
      {\}}{{{\color{delim}{\}}}}}{1}
      {[}{{{\color{delim}{[}}}}{1}
      {]}{{{\color{delim}{]}}}}{1},
}

\colorlet{magenta}{black}
\colorlet{red}{black}

\begin{document}
\title{\sys{}: Ethereum Smart COntRacTs Vulnerability Detection using Deep Neural Network and Transfer Learning}

\ifanonymous
\author{}
\else
%Oliver
\author{Oliver Lutz}
\affiliation{% (
	\country{}
\institution{University of Würzburg}
%\country{Germany}\city{Würzburg}\state{Bavaria}
}
%\email{oliver.lutz1@stud-mail.uni-wuerzburg.de}

%Huili
\author{Huili Chen}
\affiliation{%
	\country{}
\institution{University of California, San-Diego}
}
%\email{huc044@eng.ucsd.edu}

%Hossein
\author{Hossein Fereidooni}
\affiliation{%
	\country{}
\institution{Technical University of Darmstadt}
}
%\email{hossein.fereidooni@trust.tu-darmstadt.de}

%Christoph
\author{Christoph Sendner}
\affiliation{%
	\country{}
\institution{University of Würzburg}
%\country{Germany}\city{Würzburg}\state{Bavaria}
}
%\email{christoph.sendner@uni-wuerzburg.de}

%Alexandra
\author{Alexandra Dmitrienko}
\affiliation{%
	\country{}
\institution{University of Würzburg}
%\country{Germany}\city{Würzburg}\state{Bavaria}
}
%\email{alexandra.dmitrienko@uni-wuerzburg.de}

%
\author{Ahmad Reza Sadeghi}
\affiliation{%
\country{}
\institution{Technical University of Darmstadt}
}
%\email{ahmad.sadeghi@trust.tu-darmstadt.de}

%
\author{Farinaz Koushanfar}
\affiliation{%
	\country{}
\institution{University of California, San Diego}
}
%\email{farinaz@ucsd.edu}
\fi

%\maketitle
% force to have page numbers
\thispagestyle{plain}
\pagestyle{plain}

%\vspace{-1.2em}
\begin{abstract} 
Ethereum smart contracts are automated decentralized applications on the blockchain that describe the terms of the agreement between buyers and sellers, reducing the need for trusted intermediaries and arbitration. However, the deployment of smart contracts introduces new attack vectors into the cryptocurrency systems. In particular, programming flaws in smart contracts can be and have already been exploited to gain enormous financial profits. It is thus an emerging yet crucial issue to detect vulnerabilities of different classes (e.g., reentrancy or multiple send bugs) in contracts in an effective and efficient manner.
Existing machine learning-based vulnerability detection methods are limited and only inspect whether the smart contract is vulnerable, or train individual classifiers for each specific vulnerability, or demonstrate multi-class vulnerability detection without extensibility consideration.
To overcome the scalability and generalization limitations of existing works, we propose \sys{}, the first Deep Neural Network (DNN)-based vulnerability 
detection framework for Ethereum smart contracts that supports lightweight \textit{transfer learning} on unseen security vulnerabilities, thus is \textit{extensible} and \textit{generalizable}. 
\sys{} leverages a \textit{multi-output} neural network architecture that consists of two parts: (i)~A common feature extractor that learns the semantics of the input smart contract; (ii)~Multiple branch structures where each branch learns a specific vulnerability type based on features obtained from the feature extractor. 
We perform a comprehensive evaluation of \sys{} on various smart contracts.
Experimental results show that \sys{} achieves an average F1 score of \textcolor{red}{$95\%$} on six vulnerability types and the detection time is $0.02$ seconds per contract. When extended to new vulnerability types, \sys{} yields an average F1 score of \textcolor{red}{$93\%$}.
To the best of our knowledge, \sys{} is the first framework that enables transfer learning on new vulnerability types with minimal modification of the DNN model architecture and re-training overhead. 

\end{abstract}

\keywords{Blockchain, Ethereum Smart Contracts, Vulnerability Detection, Deep Neural Network, Transfer Learning}

\maketitle

\section{Introduction} 
\label{ch:introduction}
The success of Bitcoin~\cite{bitcoin} fueled the interest in the cryptocurrency platforms.
As a result, next generation blockchain-powered application platforms emerged, such as Ethereum~\cite{ethereum-whitepaper} and Hyperledger~\cite{Hyperledger}. Beyond financial transactions, these platforms provide smart contracts that are automated decentralized applications describing the terms of agreements and the transaction flow between the buyers and the sellers.

Generally, blockchains are append-only distributed and replicated databases, which maintain an ever-growing list of immutable and tamper-resistant data records (e.g., financial transactions and smart contracts). Immutability and tamper-resistance of blockchains stem from their append-only property and are paramount to the security of blockchain applications. In the context of digital currency and payments, they ensure that all the involved parties have access to a single history of payment transactions in a distributed setting and that such a history cannot be manipulated. In smart contract systems, immutability and tamper-resistance properties enforce "code is law" principle, meaning that conditions recorded in a smart contract are not to be modified since they have been written and published.

However, these properties bear their own security risks and challenges: 
First, smart contracts are written in error-prone programming languages such as Solidity~\cite{solidity}, and can contain exploitable programming errors/bugs that are often overlooked or detected only after deployment on the blockchain, and cannot be simply fixed.  
Second, Ethereum operates on open networks where everyone can join without trusted third parties while smart contracts are often in control of significant financial assets. Hence, smart contracts are attractive and easy attack targets for adversaries to gain financial profits~\cite{SmartContractsBugs}. The consequences of bug exploitation may have global effects on the entire underlying blockchain platform, far beyond the boundaries of individual contracts.
For instance, vulnerabilities in a single smart contract, the DAO~\cite{TheDAO}, affected the entire Ethereum network, when in June 2016 the attacker exploited a reentrancy bug and had withdrawn most of its funds worth about 60 million US dollars~\cite{DAO_01,DAO_02}. In the aftermath, the value of Ether, Ethereum's cryptocurrency, dropped dramatically \cite{EtherPriceDrop2016}, and the postulated "code is law" principle was undermined through the deployment of a hard fork -- a manual intervention orchestrated by a notable minority, the team of Ethereum core developers.
The Ethereum blockchain was thus split into two versions, Ethereum and Ethereum Classic, which are maintained in parallel since then. 
In another case, a critical bug accidentally triggered in 2017 resulted in the freezing of more than \$280M worth of Ether in the Parity multisig wallet~\cite{Parity2017}. 

Once it was understood that real-world attacks and even innocent mistakes can lead to fatal economic lose, the problem of detecting smart contract vulnerabilities became very appealing to the research community. 
A wide range of automated tools were developed to help find vulnerabilities in smart contracts using various techniques, such as symbolic execution~\cite{rw_securify_paper,ethereum-background-03}, satisfiability modulo theories (SMT) solving~\cite{alt2018smt}, data flow analysis~\cite{feist2019slither}, runtime monitoring~\cite{cook2017dappguard}, and fuzzing~\cite{fu2019evmfuzzer}, to name a few examples.
However, present methods provide a limited trade-off between detection effectiveness and efficiency. 
For instance, symbolic execution-based vulnerability detection is slow since all program paths need to be examined~\cite{rw_oyente_repo}, whereas data flow analysis has limited coverage~\cite{feist2019slither}. Moreover, individual tools normally cover a limited set of vulnerabilities, hence to achieve sound testing, one would normally need to apply several tools.

Recently, Machine Learning (ML) has attracted the attention of security researchers due to its capability to learn the hidden representation from the abundant data~\cite{alpaydin2020introduction}. 
Prior works have shown the effectiveness of ML techniques for detecting vulnerabilities in software written in C and C++ languages~\cite{russell2018automated,vytovtov21prediction}. 
There are also first attempts to apply ML techniques for the detection of smart contact vulnerabilities~\cite{rw_contractward,rw_lstm,huang2018hunting}. 
However, as we elaborate in details in~ Section~\ref{sec:related}, the existing solutions suffer from either of the following shortcomings: (i) They are inherently \textit{unscalable} and \textit{inflexible}, as the inclusion of any new vulnerability types would require training of new models; or (ii) They only distinguish between vulnerable and non-vulnerable smart contracts (i.e., binary classification), without the ability to detect vulnerability types; or (iii) The tools require source code of the smart contract, which limits their applicability scope.

\vspace{0.1em}
\noindent\textbf{Our goal and contributions.} In this paper, we aim to address the deficiencies and limitations of existing solutions and propose \sys{}, the first Deep Neural Network (DNN) based framework for smart contract vulnerability detection that has the following properties: (i) Operates on bytecode of smart contracts and does not require access to the source code; (ii)~ Distinguishes safe contracts from vulnerable ones with one or more vulnerabilities; 
(iii) Automatically identifies the vulnerability types of detected vulnerabilities; and 
(iv) Presents an innovative multi-output model architecture that enables fast model extension to new vulnerabilities 
using the concept of \textit{transfer learning}. 

\vspace{0.2em} 
In particular, we make the following contributions:

\begin{itemize}
    \item \sys{} enables \textit{efficient} and \textit{scalable} multi-vulnerability detection of smart contracts. It employs a \textit{multi-output} architecture where the feature extractor
    learns the program semantics and each branch of the DNN captures the semantics of a specific vulnerability class. The defender only needs to train a single DNN based on smart contract bytecode to detect multiple vulnerability types.

    \item \sys{} is the first DNN-based
    framework that supports lightweight transfer learning on new vulnerability types, thus is \textit{extensible} and \textit{generalizable}. Our multi-output architecture can be easily expanded by concatenating a new classification branch to the feature extractor. Only the new branch needs to be trained on the new dataset during transfer learning. 
    
    \vspace{0.1em}
    \item \sys{} is automated for inspecting vulnerabilities in smart contracts prior to their deployment and demonstrates superior vulnerability detection performance while incurring low overhead. We perform a comprehensive evaluation of \sys{} across various Ethereum smart contracts to corroborate its effectiveness, efficiency, and extensibility. Our framework can detect vulnerabilities in $0.02$ seconds per smart contract and yields an average F1 score of 95\% across all evaluated classes.  

    \vspace{0.1em}
    \item We evaluate \sys{} using a dataset consisting of 93.497 smart contracts downloaded from Ethereum blockchain and labeled them using three exiting vulnerability detection tools: Oyente~\cite{rw_oyente_repo}, Mythril~\cite{rw_mythril_repo}, and Dedaub~\cite{dedaub}. To construct and label such a dataset, we developed a toolchain \datatool{} that automates the acquisition and labeling process. \datatool{}'s modular structure enables one to easily integrate other tools for labeling. We will open source \datatool{} and our dataset to encourage further research on smart contract security.   
\end{itemize}

Overall, \sys{} is the first flexible and generalizable smart contract detection technique with superior vulnerability detection performance.
\vspace{0.1em}
\noindent\textbf{Outline.} The remaining part of the paper is organized as follows: In Section~\ref{sec:background}, we provide the necessary background information on smart contracts and their vulnerabilities, as well as give insights into deep learning. 
We define the threat model and identify the design challenges in Section~\ref{sec:overview}.
Section~\ref{sec:design} sheds light into \sys{}'s design specifics, while Section~\ref{sec:implementation} and~\ref{sec:evaluation} provide implementation details and evaluation results, respectively. After that, the relevant related work is surveyed in Section~\ref{sec:related}. We finally conclude in Section~\ref{sec:conclusion}. 
 \vspace{-0.5em}
\section{Background}
\label{sec:background}
We introduce smart contracts and their vulnerabilities in Section~\ref{sec:smart_contracts}, and the background on deep learning (DL) in Section~\ref{sec:dl}.

 \vspace{-0.5em}
\subsection{Smart Contracts and Vulnerabilities}
\label{sec:smart_contracts}
\vspace{-0.2em}
A smart contract is written in high-level programming languages such as Solidity~\cite{solidity} and is called by its address to run operations on the blockchain.
Once compiled, the bytecode of the contract is generated and executed inside the Ethereum Virtual Machines (EVM). 
Since there is a one-to-one mapping between a blockchain operation and bytecode representation, it is feasible to analyze the control flow of a contract at the bytecode-level. When triggered, the execution of the smart contract is autonomous and enforceable for all participating parties~\cite{ethereum-background-04}. 

The EVM itself is a stack-based machine with a word size of 256 bits and stack size of 1024~\cite{ethereum-yellowpaper}. The memory applies a word-addressable model. Once a contract is deployed on the blockchain, it requires gas to function. Gas is the unit used to pay the computational cost of the miners running contracts or transactions and is paid in Ether.

Similar to any other software, smart contracts might suffer from vulnerabilities and programming bugs. 
The Smart Contract Weakness Classification (SWC) Registry~\cite{swcregistry} collects information about various vulnerabilities. 
We differentiate five categories of vulnerability types: External Calls, Programming Errors, Execution Cost, Influence by Miners, and Privacy. 

\textbf{External Calls}. Any public function of a smart contract can be called by any other contract. A malicious user can then exploit public availability to attack vulnerable functions of smart contracts. A prominent example is the so-called reentrancy bug (SWC-107~\cite{swcregistry}). Here, an attacker can call a contract's function multiple times before the initial call is terminated. If the internal contract state is not securely updated, the attacker can drain Ether from the contract by recursively calling the function.

\textbf{Programming Errors}.
Some of the programming errors in smart contracts are very similar to those found in traditional programs,
such as missing input validation, typecast bugs, use of untrusted inputs in security operations, unhandled exception, exception disorder, and Integer overflow and underflow vulnerabilities. In another example, an \textit{assert} function used in tests and not removed by the programmer in the release version may lead to its misuse by an attacker, which can result in exploitable error conditions (SWC-110~\cite{swcregistry}). Other vulnerabilities can be specific to smart contracts.

Examples are greedy contracts that lock Ether indefinitely, gasless send bug that does not provide sufficient gas to execute the code of the smart contract, Ether lost in transfer if sent to unknown recipients, etc. 
Further examples are \textit{callstack depth} 
limit reached exception bug and unprotected \textit{selfdestruct} instructions (see SWC-106~\cite{swcregistry}), where an
attacker can call a smart contract's public function containing a \textit{selfdestruct} to terminate the smart contract, or he can fill up the stack to reach the stack size limit. Both attacks result in a Denial of Service (DoS) of vulnerable smart contracts.

\textbf{Execution Cost}. Every transaction on the Ethereum network costs gas. However, every block has a spendable gas limitation. An attacker can use this limit to induce a DoS of a vulnerable contract. For example, if the execution time of a function is dependent on input from the caller, a malicious caller can expand the execution time of the smart contract over the gas limit (SWC-128~\cite{swcregistry}). Thereby, execution is terminated by exceeding the gas limit before it is finished. Another way an attacker can misuse the gas limit per block is to induce an error on a \textit{send} call. If a programmer bundles \textit{multiple sends} in one function of the smart contract, the attacker can then prevent the execution of other \textit{send} calls in the function.

\textbf{Influence by Miners}. Miners are entities that actually execute transactions on the blockchain. They can decide which transactions to execute, in what order, and also able to influence environment variables (e.g., timestamps). 
To illustrate the problem, let us assume a scenario where a smart contract is instructed to send Ether to the first user that solved the puzzle. If two users commit a transaction with the solution at the same time, a miner decides who will be first and therefore will be getting the Ether (SWC-114~\cite{swcregistry}). This vulnerability type is generally referred to as Transaction Order Dependence (TOD).

\textbf{Privacy}. Solidity~\cite{solidity} offers different visibility labels for functions and variables. Most notably, a programmer can define a function as private or public. The default setting for functions is public, which can be overlooked by the programmer (SWC-100~\cite{swcregistry}). If a variable or function is set to private, it can't be seen or changed by other contracts. However, even if it is set as private, an attacker can still parse the blockchain data, where those variables are stored in plaintext (SWC-136~\cite{swcregistry}). 

\begin{table*}
\centering
\scalebox{0.93}{
\begin{tabular}{|c|c|c|c|c|c|c|} 
\hline
\textbf{Name}  & \textbf{Method}  & \textbf{Capability}  & \textbf{Extensible}  & \begin{tabular}[c]{@{}c@{}}\textbf{Required }\\\textbf{ Input} \end{tabular} & \begin{tabular}[c]{@{}c@{}}\textbf{Detection time }\\\textbf{ per Contract (s)} \end{tabular} & \textbf{F1-score}  \\ 
\hline
\textbf{This work}  & \begin{tabular}[c]{@{}c@{}}ML (LSTM) \\ + transfer learning \end{tabular} & Multi-label & Yes & Bytecode & 0.02 & \textcolor{red}{0.95} \\ 
\hline
Oyente~\cite{ethereum-background-03} & Symbolic execution & Multi-class & No & \begin{tabular}[c]{@{}c@{}}Source code\\ and bytecode \end{tabular} & 350 & 0.38 \\ 
\hline
Mythril~\cite{rw_mythx_article} & \begin{tabular}[c]{@{}c@{}}Symbolic execution,\\ taint analysis, and SMT \end{tabular} & Multi-class & No & Bytecode & 11.1 & 0.47 \\ 
\hline
Dedaub~\cite{rw_dedaub_contract_library_page} & Flow and loop analysis & Gas-focused vulnerability & No & Source code & 20 & NA \\ 
\hline
Securify~\cite{rw_securify_paper} & Symbolic analysis & Binary decision & No & Bytecode & 30 & 0.54 \\ 
\hline
Vandal~\cite{brent2018vandal}  & \begin{tabular}[c]{@{}c@{}}Logic-driven \\ static program analysis \end{tabular} & Multi-class & No & Bytecode & 4.15 & NA \\ 
\hline
\multicolumn{1}{|l|}{ContractWard~\cite{rw_contractward}} & ML (bigram model) & Binary decision & No & Opcode & 4 & 0.96 \\ 
\hline
Towards Sequential~\cite{rw_lstm} & ML (LSTM) & Binary decision & No & Opcode & 2.2 & 0.86 \\ 
\hline
NLP-inspried~\cite{gogineni2020multi} & ML (AWD-LSTM) & Multi-class & No & Opcode & NA & 0.9 \\ 
\hline
Color-inspried~\cite{huang2018hunting} & ML (CNN) & Multi-label & No & Bytecode & 1.5 & 0.94 \\ 
\hline
Graph NN-based~\cite{zhuangsmart} & ML (GNN) & Multi-class & No & Source code & NA & 0.77 \\
\hline
\end{tabular}
}
\vspace{0.3em}
\caption{Qualitative comparison of \sys{} and existing smart contract vulnerability detection methods. \label{tab:comparison}}
\vspace{-1em}
\end{table*}

\vspace{-0.3em}
\subsection{Deep Learning}
\label{sec:dl}
\vspace{-0.2em}
\sys{} operates on the bytecode representation of the smart contracts, which can be considered as a special case of text data. 
We introduce background about text representation and recurrent neural networks below.

\vspace{0.1em}
\textbf{Text Representation.} 
The text modality is typically transformed into numerical vectors for usage in 
ML algorithms. 
This transformation can be realized in different ways, such as a bag of words~\cite{wallach2006topic}, n-gram language model~\cite{brown1992class}, and embedding layer~\cite{zamani2017relevance}.
The numerical vectors converted from the text data are then used as the direct input to the DL models. 

%% RNN hidden states 
% \vspace{-0.5em}
\setlength{\belowcaptionskip}{-10pt}
\begin{figure}[t!]
  \centering
  \includegraphics[width=0.86\columnwidth]{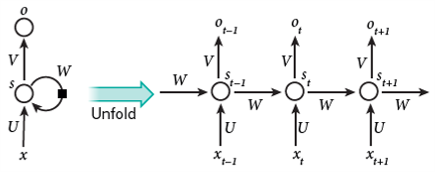}
   \vspace{-0.7em}
  \caption{A recurrent neural network and its unfolding in time~\cite{deep-learning-background-03}. The computation of the forward pass is shown.}
  \label{fig:rnn}
  \vspace{-0.3em}
\end{figure}
\vspace{0.1em}
\textbf{Recurrent Neural Network.}
An RNN is a category of DNNs where the connections between nodes (i.e., neurons) construct a direct computational graph along the temporal sequence~\cite{kim2017residual}. 
A key property of the RNN is that the model can use its internal states as memory cells to store the knowledge about prior inputs, thus capturing the contextual information in the sequential input data (e.g., text document).  
As shown in Figure~\ref{fig:rnn}, the hidden states obtained from the previous input ($s_t$) affects the output in the current time step ($o_{t+1}$). The unfolded RNN diagram reveals the \textit{`parameter sharing'} mechanism of RNNs where the weight matrices ($W$, $V$, and $U$) are shared across different time steps. Parameter sharing makes RNNs generalizable to unseen sequences of different lengths.   

\textbf{Multi-label vs. Multi-class classification.} 
Smart contracts vulnerability detection can be realized with two different paradigms. The first one is known as  
\textit{Multi-class classification}, which refers to the case where the classification task has more than two classes that are mutually \textit{exclusive}. In particular, each sample is assigned with {one and only one} class label.
The second one, \textit{Multi-label classification}, also involves multiple classes while a data sample can have more than one associated labels. This is because the classes in multi-label tasks describe non-exclusive attributes of the input (e.g., color and length).

\begin{figure}[ht!]
  \centering
  \includegraphics[width=0.9\columnwidth]{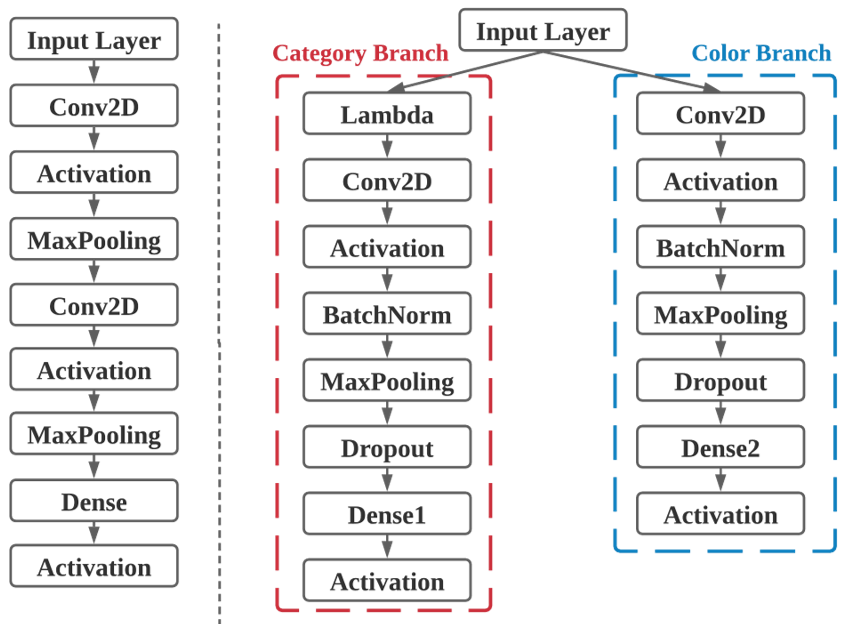}
  \vspace{-0.5em}
  \caption{DNN architectures of multi-class (left) and multi-label (right) formulation of the clothing classification task~\cite{multi_output}. 
  }
  \label{fig:multioutput_nn}
  \vspace{-0.3em}
\end{figure}

Let us use an example to illustrate the difference between these two paradigms.  
Given a clothing dataset with three colors (black, blue, red) and four categories (jeans, dress, shirt, shoes), we want to train a model to predict these two clothing attributes simultaneously. 
Figure~\ref{fig:multioutput_nn} shows the architecture of the multi-class and multi-label formulation of the clothing classification task. 
The multi-class design has only one set of dense layers (i.e., `heads') at the bottom of the DNN where the last dense layer has $3\times4=12$ neurons.
The network topology for multi-label classification has two sets of dense heads at the end of the DNN where the last dense layer in each branch has $4$ and $3$ neurons to learn the clothing category and color attribute, respectively.  
We call the network design of the \textit{multi-label} classification with multiple sets of dense heads as \textit{`multi-output'} architecture throughout this paper.  
The `stem-branch' topology makes the multi-output architecture extensible to learn new attributes. 
\sys{} leverages this observation to devise an efficient and extensible smart contract inspection solution. 

\vspace{-1em}

% \vspace{-0.5em}
\section{Motivation} \label{sec:motiv}

Vulnerability detection of smart contracts is a challenging task.  
Table~\ref{tab:comparison} summarizes the properties and performance metrics of previous detection tools. 
To the best of our knowledge, none of the existing works takes into account of the extensibility requirement of vulnerability detection in their design. This means that when new vulnerabilities of smart contracts are discovered and exploited by the adversary, the detection tools cannot quickly adapt to the new contracts. 
\sys{} is motivated to provide an efficient and extensible solution to concurrent detection of multiple vulnerability types using DL techniques.
To this end, we first investigate two research questions (RQs) about smart contract inspection in this section and answer them in details in Section~\ref{sec:design}. 

\vspace{0.2em}
\textit{RQ1: How to build a single DL model for detecting multiple vulnerability types?}

Prior works have attempted to use ML algorithms to inspect the vulnerabilities in smart contracts.  
An LSTM-based detection approach is proposed in~\cite{rw_lstm} to distinguish vulnerable contracts from safe ones (i.e., binary classification), which does not meet the goal of RQ1. 
ContractWard~\cite{rw_contractward} uses `One vs. Rest' algorithms and trains individual ML models for each vulnerability class. 
Therefore, ContractWard~\cite{rw_contractward} does not meet the model number constraints in RQ1.  
The paper~\cite{huang2018hunting} suggests to transform contract bytecode to RGB images and employs a Convolutional Neural Network (CNN) for binary classification.
Their method can support multi-label classification by re-training the obtained CNN on the corresponding dataset, thus satisfying both requirements in RQ1.

\vspace{0.3em}
\textit{RQ2: How to make the DL model extensible to new vulnerabilities?}

ContractWard~\cite{rw_contractward} can be adapted to detect a new vulnerability type by training a \textit{new ML model} on the n-gram features from scratch.
The color-inspired inspection method~\cite{huang2018hunting} does not consider the extensibility requirement of the detection. 
However, if applying the traditional transfer learning approach in the ML domain, the defender may replace the last dense layer of the pre-trained CNN with a new one. The new layer is then trained on the mixture of the old vulnerability data and the new one.  
Note that training a new ML model from scratch or fine-tuning a pre-trained ML model on a large dataset incurs high computational cost, thus is not desirable in practice. 

\vspace{-0.8em}
%\section{Overview}  \label{sec:overview}
\vspace{-0.3em}
\section{Threat Model and Challenges}  \label{sec:overview}
\vspace{-0.3em}
We introduce our threat model in Section~\ref{sec:threat} and the challenges of developing an effective vulnerability detection technique for smart contracts in Section~\ref{sec:challenges}.

\vspace{-0.5em}
\subsection{Threat Model} 
\label{sec:threat}
\vspace{-0.2em}
We consider a scenario where the attacker is a malicious party that can obtain knowledge from any public data structure in the blockchain and can upload his contract code to the Ethereum system. The exact attack requirements and actions that the adversary need to take for the exploitation of a smart contract are specific to a vulnerability class, which we explain in Section~\ref{sec:evaluation:dataset}.
Note that \sys{} is generic and we demonstrate its effectiveness with eight common vulnerability types in this paper. 
The defender can be an Ethereum contract designer that aims to ensure the program is not exploitable by malicious adversaries during the code development time. This role can also be taken by the end user that intends to verify the security of the contract at runtime before sending any transactions to it.
\textcolor{magenta}{We assume the performance metrics reported in the previous papers~\cite{ethereum-background-03,mueller2018smashing,rw_dedaub_mad_max_paper} as well as the open-sourced implementation of existing detection tools~\cite{rw_oyente_repo,rw_mythril_repo,dedaub} are reliable. This assumption is feasible since expert inspection has been performed to cross-validate the performance of proposed detection methods in the previous works. \sys{} relies on this assumption since our method is an instantiation of supervised learning paradigm.}

\sys{} aims to provide the defenders with a holistic smart contract vulnerability detection solution. To this end, we formulate vulnerability inspection as a
\textit{supervised} \textit{multi-label classification} problem where the input is the contract (can be represented in high-level language, opcodes, or bytecodes) and the output is the corresponding vulnerability types as introduced in Section~\ref{sec:evaluation:dataset}.

\setlength{\belowcaptionskip}{-12pt}
\begin{figure*}
  \centering
  \includegraphics[width=0.83\textwidth]{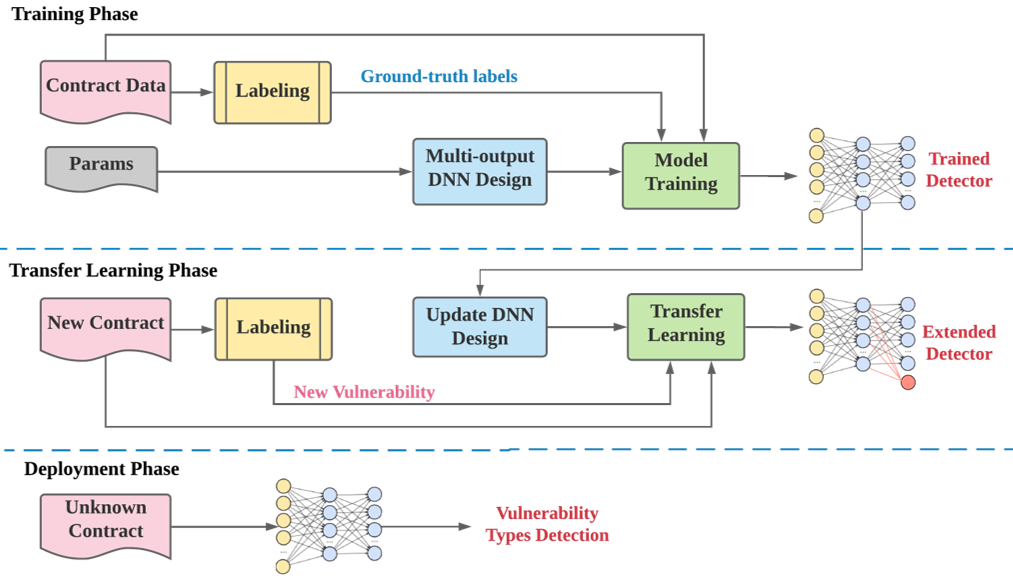}
  \vspace{-0.5em}
  \caption{Overview of \sys{} framework for Ethereum smart contract vulnerability detection. \sys{} has three stages: training, transfer learning, and deployment as shown in the top, middle, and bottom part of the figure. 
  }
  \label{fig:global}
  \setlength{\belowcaptionskip}{-12pt}
\end{figure*}

\vspace{-0.5em}
\subsection{Challenges} \label{sec:challenges}
We identify the challenges to develop an effective DNN for automated smart contract
vulnerability detection below. 

\textbf{(C1) Data Collection and Pre-processing.} Supervised learning of a DNN requires a sufficiently large labeled dataset. Analyzing smart contracts using the source code is difficult since not that many smart contracts are open-sourced. On another hand, blockchain platforms typically host their smart contracts in a form of a bytecode, which is publicly available.
However, performing analysis on the bytecode raises another challenge due to the large bytecode length of long contracts.
This long sequence requires a large memory footprint during model training, suggesting that using the bytecode as direct input to the DNN model is impractical. 

\vspace{0.1em}
\textbf{(C2) Feature Extraction.} 
The second challenge concerns the problem of finding the proper feature representation of the smart contract program. 
On the one hand, manual design of contract features is time-consuming and has limited efficacy, since the contract bytecode is long and hard to interpret by human developers. On the other hand, current implementations of automated feature extraction~\cite{rw_contractward,rw_lstm,zhuangsmart}

(Section~\ref{sec:related}) mainly apply traditional software testing techniques on the smart contract without exploring its domain-specific properties, thus taking a long time to inspect a single contract.

\vspace{0.1em}
\textbf{(C3) Dataset Imbalance.} 
The third challenge is a class imbalance in the training set. 
Prior works have shown that the number of contracts with specific vulnerability types is much lower than the one of non-vulnerable contracts~\cite{rw_contractward}.
Learning the characteristics of vulnerable contracts with a DNN is challenging since the stochastic gradient descent (SGD) based learning of DNN models is inherently biased towards the majority class, while our objective is to recognize the minority class (i.e., vulnerable contracts). As such, it is crucial to provide the DNN model with sufficient vulnerable contracts to ensure a high true positive rate.

\vspace{0.1em}
\textbf{(C4) Efficiency.} 
The fourth challenge is to ensure the efficiency of DNN training and inference for concurrent detection of multiple vulnerability types. Identifying diverse attacks with a single DNN detector is challenging since different vulnerabilities exploit distinct loopholes in the contract, which might be hard to capture with a conventional DNN.
Efficiency is important for practical development of the contract scanner since devising individual classifiers for each vulnerability class, as done, e.g., in~\cite{rw_contractward}, is unscalable and incurs large computation overhead. 

\vspace{0.2em}
\textbf{(C5) Extensibility/Generalizability.} 
Finally, smart contract inspection should be capable of learning new vulnerability types quickly while preserving the knowledge of the known ones. We call this requirement `extensibility'. 
This property is important for both, researchers and practitioners, since new attacks on smart contracts are emerging at a fast speed.
Augmenting an existing contract detector to new attacks is non-trivial since the new attack exploits the unseen and unpredictable susceptibilities of smart contracts compared to the previously known attacks.
Training a new DNN from scratch to accommodate the new vulnerabilities consumes extensive resources and incurs additional engineering costs. 

We show how \sys{} framework tackles each of the above challenges in Section~\ref{sec:design}. 

\vspace{-0.1em}
\section{\sys{} Design}
\label{sec:design}
We propose \sys{}, the first {extensible} and {transfer learning-friendly} DNN-based framework for vulnerability detection of Ethereum smart contracts. 
The key innovation of \sys{} is that we {decompose} the task of vulnerability detection into two subtasks: (T1) Learning the bytecode features of general contracts (\textit{attack-agnostic}); (T2) Learning to identify each particular vulnerability class (\textit{attack-specific}). To achieve (T1), we design a \textit{feature extractor} that captures the semantic and syntactic information of contract bytecode regardless of its vulnerabilities. To perform (T2), we devise an individual \textit{vulnerability class branch} to characterize susceptibility given the bytecode features extracted in (T1). Our \textbf{divide-and-conquer} design is highly modular and flexible compared to previous detection techniques as we corroborate in Section~\ref{sec:evaluation}. 

Below we discuss each step of \sys{} design, which aims to overcome the design challenges in Section~\ref{sec:challenges}. 

\vspace{-0.3em}
\subsection{\sys{} Global Flow} \label{sec:global}

Figure~\ref{fig:global} demonstrates the global overview of \sys{} smart contract detection technique. Our framework has three stages: DNN classifier training, transfer learning, and deployment. We discuss each stage below.

\vspace{0.05em}
\textbf{Training.} The top part of Figure~\ref{fig:global} shows \sys{}'s training pipeline. 
To enable supervised learning for vulnerability detection, the defender first constructs the smart contracts bytecode dataset with corresponding labels, as detailed in Section~\ref{sec:data}.
The defender specifies the system parameters, including the vulnerabilities of interests and the available hardware resources for \sys{}'s multi-output DNN design.  
Finally, the devised model is trained on the collected contract data with their corresponding labels, resulting in a converged DNN detector.

\vspace{0.05em}

\textbf{Transfer Learning.} The middle part of Figure~\ref{fig:global} shows the transfer learning phase of \sys{}. 
Given a trained detector and the bytecode of smart contracts with new vulnerabilities, \sys{} extends the DNN architecture devised in the original training phase with new parallel branches. 
The layers in the expanded branch are then trained on the new vulnerability data with the associated labels to perform transfer learning. 
To the best of our knowledge, \sys{} is the first framework that supports transfer learning to accommodate new vulnerability types of the smart contracts.

\vspace{0.05em}
\textbf{Deployment.} The bottom part of Figure~\ref{fig:global} shows the workflow of \sys{}'s deployment stage. 
After the training/transfer learning phase completes, \sys{} returns a trained DNN classifier that can detect whether an unknown smart contract has any of the learned vulnerability types.  

\vspace{-0.6em}
\subsection{Neural Network Design}
\label{sec:approach:deep-learning}
\vspace{-0.3em}
Prior ML-based vulnerability detection techniques only explored simple architectures, such as K-Nearest Neighbors, SVM, Decision Trees~\cite{rw_contractward}, or use a CNN~\cite{huang2018hunting}/RNN~\cite{rw_lstm} to decide whether a smart contract is vulnerable or not. 
To overcome these constraints, \sys{} aims to design an \textbf{extensible} DNN detector that: (i) Provides the probability that the smart contract has certain vulnerabilities, instead of making a binary decision about contract security; (ii) Classifies multiple vulnerabilities with a single DNN. 
To this end, we propose an innovative \textbf{multi-output architecture} for \textbf{concurrent} detection of multiple vulnerability types. 
This neural network design step is shown by the `Multi-output DNN Design' module in \sys{}'s global flow (Figure~\ref{fig:global}). 

Figure~\ref{fig:demo_nn} illustrates the generic architecture of our DNN detector. {The stem and branch layers shown in the diagram are typical DNN layers such as the Dense layer, Dropout layer, and GRU layer.}
The multi-output model has two main components and we discuss each one below:  

%% --- our NN ---- %% 
% \vspace{-0.8em}
\begin{figure}[ht!]
  \centering
  \includegraphics[width=0.98\columnwidth]{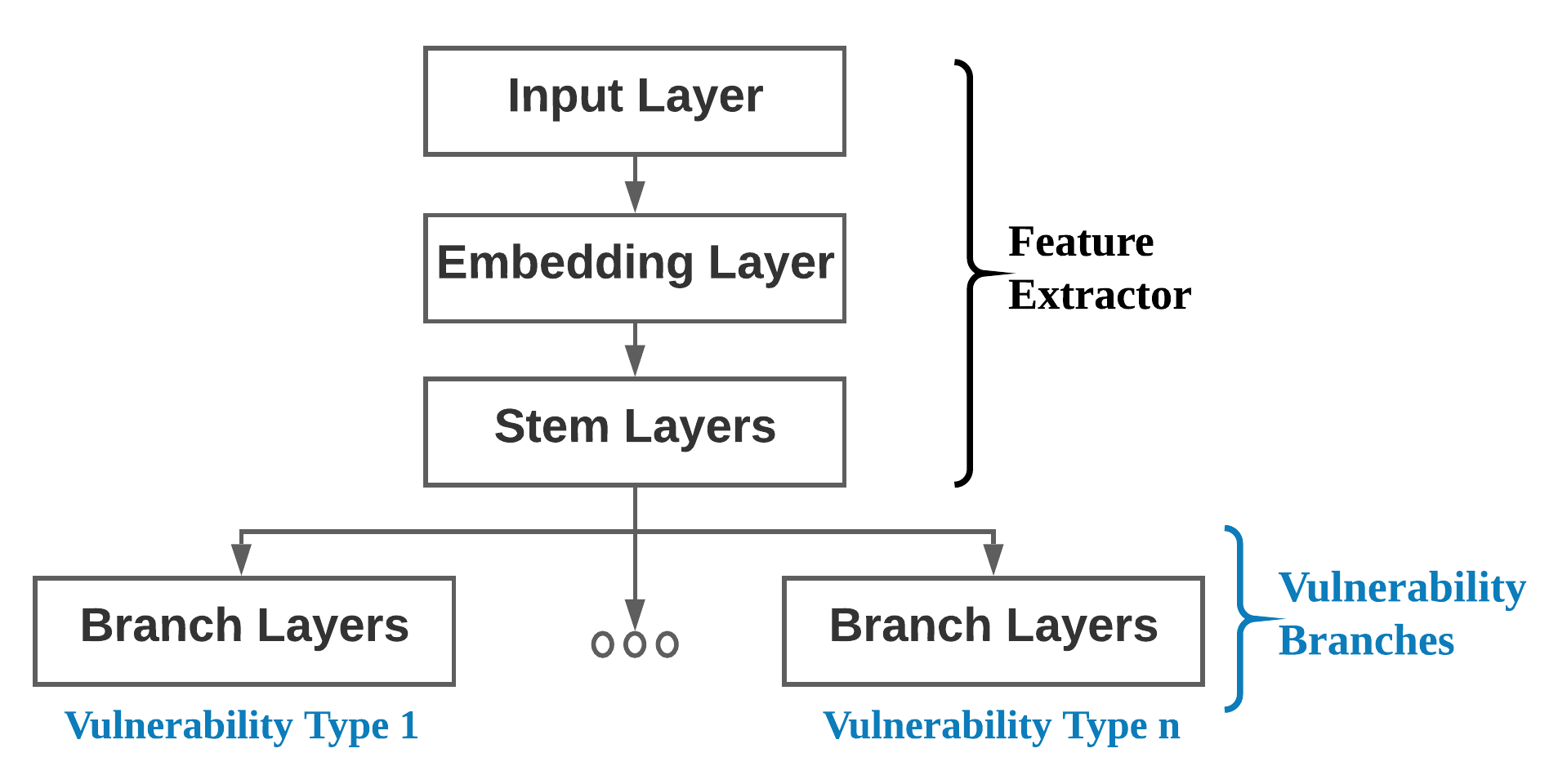}
  \vspace{-0.6em}
  \caption{General architecture of \sys{}'s multi-output model for concurrent detection of multiple vulnerability types. 
  }
  \label{fig:demo_nn}
  \vspace{0.3em}
\end{figure}

\textbf{(i) Feature Extractor.} The first component of \sys{}'s extensible DNN model is the common feature extractor (i.e., `stem') shared by all the bottom branches. 
The feature extractor is a stack of layers that learn the fundamental features in the input data that are general and useful across different attributes. 
In the context of smart contracts, the feature extractor is trained to learn the semantic and syntactic information from the contracts' bytecode. 
To this end, we incorporate several key layers in \sys{}'s feature extractor: 

\begin{itemize}
    \item \textit{Embedding Layer.} The bytecode of smart contracts are long hexadecimal numbers, while DNNs typically work with fractional numbers to achieve high accuracy. The embedding can solve this discrepancy since it stores the word embedding in the numerical space and retrieves them using indices~\cite{zamani2017relevance}.
    The embedding layer provides two key benefits: (i)~Compressing the input via a linear mapping, thus reducing the feature dimension; (ii)~Learning bytecode in the embedding space (fractional numbers). This facilitates representation exploration and gathers similar bytecode in the vicinity of each other. 
    \sys{} leverages the advantages of the embedding layer to capture the semantics in the input bytecode. 
    
    \vspace{0.3em}
    \item \textit{GRU/LSTM Layer.} The stem layers and branch layers in Figure~\ref{fig:demo_nn} can include GRU/LSTM layers for processing sequential inputs. 
    Gated Recurrent Units (GRU) and Long Short-Term Memory (LSTM) are two typical layers in recurrent neural networks that help to overcome the short-term memory constraint and vanishing gradient problem~\cite{kim2017residual} using a \textit{`gating'} mechanism. 
    More specifically, both types of layers have internal gates that regulate the information flow along the time sequences and decide which data shall be kept/forgotten. 
    We mainly use GRU layers in \sys{}'s DNN design. 
\end{itemize}

\textbf{(ii) Vulnerability Branches.} 
The second component of \sys{}'s multi-output DNN architecture is the \textbf{ensembling} of multiple vulnerability branches. Each branch is a stack of layers that are trained to learn the patterns/hidden representation of the corresponding vulnerability class. 
While there is no direct dependence between different branches, they share the same feature extractor, i.e., the input to each branch is the same. 
This is feasible since the branch input (which is also the feature extractor's output) shall capture the semantics in the contract's bytecode, which is common/general information useful for different vulnerabilities.

\sys{}'s `stem-branches' architecture is similar to the \textit{`mixture of experts'} paradigm where the problem space is divided into homogeneous regions and individual expert models (learners) are trained to tackle each sector~\cite{kiyono2019mixture}. 
The main difference between \sys{}'s multi-output design and the mixture of experts model is that the latter one requires a trainable gating network to decide which expert shall be used for each input region, while our DNN model does not need such a gating mechanism since we aim to detect multiple vulnerability types of the input contract in parallel.  

Note that the last layer of each vulnerability branch is a Dense layer with one neuron. 
The \textit{sigmoid} evaluation of this neuron's activation value gives the \textit{probability} that the input contract has the specific vulnerability. 
As such, \sys{} engenders detection results with better \textbf{interpretability} by providing the confidence score for its diagnosis instead of the binary decision about vulnerability existence. 

In summary, \sys{}'s multi-output architecture solves the feature extraction, efficiency and extensibility challenges (C2, C4, C5) identified in Section~\ref{sec:challenges}.
In particular, \sys{} allows the defender to train a \textit{single} DNN for detecting multiple vulnerability types instead of training an individual classifier for each attack, thus demonstrating superior efficiency compared to the prior work~\cite{rw_contractward,huang2018hunting,rw_lstm}. 
\sys{} design incurs minimal non-recurring engineering cost and is scalable as new vulnerabilities are identified.

\vspace{-0.3em}
\subsection{Transfer Learning} \label{sec:transfer}
\vspace{-0.3em}
Malicious parties have a strong incentive to discover and exploit new vulnerabilities of smart contracts due to the associated prodigious profits. 
As such, the contract inspection technique shall be extensible to learn new vulnerabilities as they are identified.  
We propose \textit{transfer learning} as the solution to the challenge (C5) in Section~\ref{sec:challenges}.
More specifically, we suggest to \textit{expand} the pre-trained multi-output DNN model by adding new vulnerability branches for transfer learning. 
This process is demonstrated in the middle part of Figure~\ref{fig:global}. 
The transfer learning capability of \sys{} ensures that our detection framework can be upgraded with the minimal cost to defend against emerging attacks on smart contracts. 

Our transfer learning stage has two goals: 

\vspace{0.1em}
\textbf{(G1) Preserve Knowledge on Old Vulnerabilities.}
On the one hand, the DNN detector shall retain knowledge about the previous vulnerability types that are used in the initial training phase. 
This property is important since \sys{} aims to provide a holistic and extensible solution to concurrent detection of multiple vulnerabilities. 
As such, maintaining high classification accuracy on the known attacks is essential. 

\vspace{0.1em}
\textbf{(G2) Learn New Vulnerabilities Quickly.} On the other hand, transfer learning aims to adapt the pre-trained model to achieve high accuracy on the new dataset in an efficient way. 
This is also required by the extensibility challenge (C4) in Section~\ref{sec:challenges}. 
To achieve fast adaptation, transfer learning shall yield minimal runtime overhead. 
This requirement is crucial for practical deployment, since training a new DNN model from scratch for the new vulnerabilities is prohibitively expensive and hard to maintain. 

\sys{}'s transfer learning phase works as follows. 
When a new vulnerability is identified, the defender constructs a new training dataset accordingly and updates the converged DNN detector by adding a new vulnerability branch (i.e., the stack of layers). 
During transfer learning, the parameters of the common feature extractor and existing vulnerability branches are \textit{fixed}. 
Only the parameters in the newly added branch are updated with the new vulnerability dataset. 
Freezing the feature extractor and the converged branches ensures that the updated DNN classifier preserves the detection accuracy on the old vulnerabilities (G1), and training a new branch enables the updated model to learn the new attack (G2).

\textcolor{magenta}{Besides extensibility, \sys{} also enables lightweight and fast adaptation when \textit{mode drift} occurs. Smart contracts running on Ethereum are dynamic and change over time, which might lead to a decrease of \sys{}'s performance. To alleviate the model drift concern, the contract developer can update the parameters of the vulnerability branches given a set of labeled new contracts while fixing the ones of the feature extractor. }
\vspace{-0.2em}
\section{Dataset Construction Toolchain}
\label{sec:data}
\vspace{-0.2em}
In this section, we present \datatool{}, the toolchain we built to construct the labeled smart contract dataset and to address the challenge (C1) defined in Section~\ref{sec:challenges}. 
Details about our dataset used in DNN training is given in Section~\ref{sec:evaluation:dataset}. 
It is worth mentioning that a standard and public smart contract vulnerability dataset was not available before. We will open source \datatool{} toolchain and our dataset used in the evaluation to facilitate the development and comparison between emerging detection techniques.

%% 
% \vspace{0.8em}
\subsection{Design Choices}
\label{sec:dataset:choices}
\vspace{-0.2em}
To build a sufficiently large training set for our supervised DNN training, we make the design choice to work on the bytecode-level since the bytecode of smart contracts are publicly available. The accessibility of contract bytecode solves the challenge (C1) in Section~\ref{sec:challenges}.
This also makes our approach agnostic to the programming languages of smart contracts, since smart contracts are written in different languages (e.g., Solidity~\cite{solidity}, Viper~\cite{vyper}, and Serpent~\cite{serpent}) get eventually compiled to the same bytecode.  
The EVM bytecodes are executed in a stack context, thus the control flow of a program at the bytecode level contains useful information for detection.

\datatool{} is designed to obtain the bytecode files of smart contracts from the Ethereum platform, \textcolor{magenta}{label them using available bytecode-level detection tool(s), }
and store the result in a database.
Note, that we concentrate on Ethereum smart contracts for exemplary purposes in this paper. Generally, \sys{} can also be used for vulnerability detection in smart contracts of other cryptocurrency platforms that use Ethereum-compatible EVM, such as Quorum~\cite{quorum}, Vechain~\cite{vechain}, Rootstock~\cite{rootstock}, and Tron~\cite{tron}, to name a few. This is possible since the bytecode of smart contracts on these platforms is compatible with Ethereum EVM. 

% \vspace{-0.3em}
\subsection{\datatool{} Architecture}
\label{sec:implementation:bytecode}

\vspace{-0.5em}
\begin{figure}[ht!]
  \centering
  \includegraphics[width=\columnwidth]{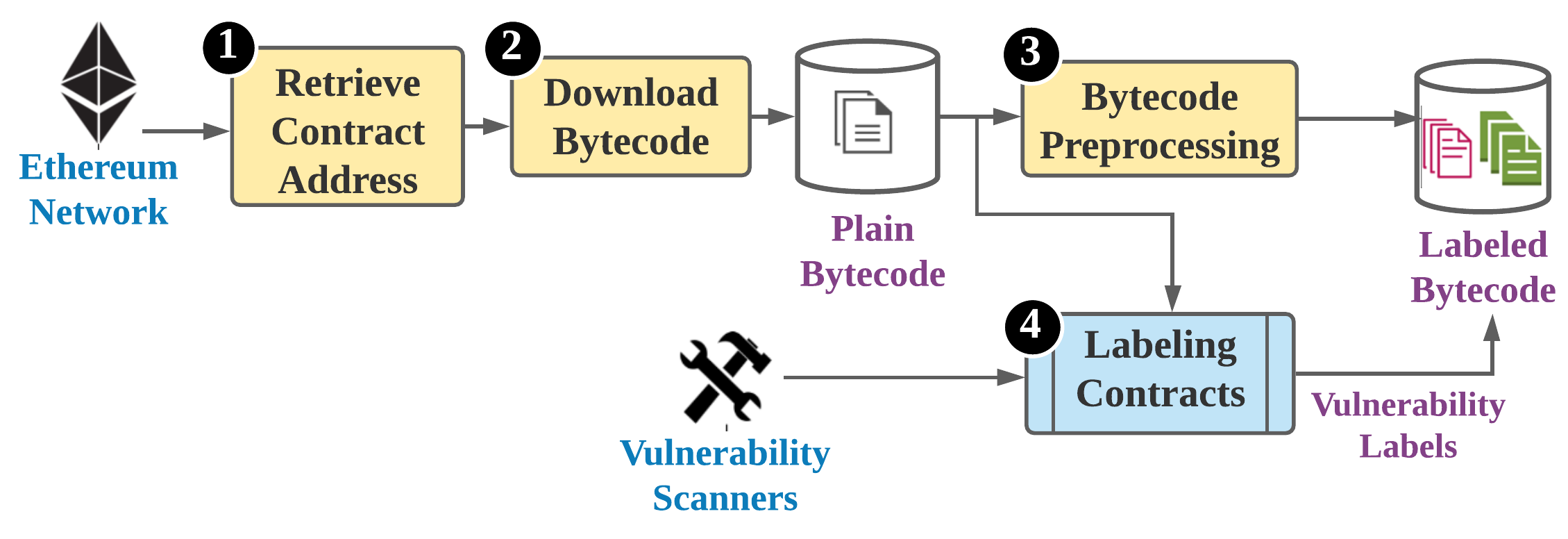}
  \vspace{-2em}
  \caption{Generic workflow of \datatool{} for smart contract acquisition and labeling.}
  \label{fig:data:toolchain}
  \vspace{0.3em}
\end{figure}

We show the generic workflow of \datatool{} toolchain in Figure~\ref{fig:data:toolchain}.
In step~(1), the addresses of contracts are retrieved from the blockchain. Step (2) involves downloading 
the bytecode from the Ethereum network by address and extracting information into a database. In step (3), the bytecode is pre-processed for input efficiency. 
\textcolor{magenta}{The last step (4) outputs the vulnerability types of the contracts using the bytecode-level detection tools. It is worth noticing that any vulnerability detection techniques that take bytecode as input can be used by \datatool{} for contract labeling, including existing methods such as Oyente~\cite{rw_oyente_repo}, Mythril~\cite{rw_mythril_repo}, Dedaub~\cite{dedaub}, and our proposed new method \sys{}. We show in Section~\ref{sec:evaluation} that \sys{} outperforms existing tools in terms of both detection effectiveness and efficiency.}

For a concrete instantiation of the generic flow, we use the open-source tool Ethereum ETL~\cite{ethereum-etl-article,ethereum-etl-repo} to retrieve the addresses of contracts in step (1). The ETL tool connects to the Ethereum network and exports blockchain content into CSV files. For step (2), we also utilize existing tools: Multiple Web APIs exist to download the contracts' bytecode by its address. We opt for two APIs for instantiating the bytecode downloading module of \datatool{}: Infura~\cite{infura} and Dedaub's Contract-Libray~\cite{dedaub}. For the instantiation of step (3), we develop an assistive Python module named \textit{Contract Loader} to extract the information from smart contracts into a MySQL database~\cite{mysql}. 
In the last step (4), we use three tools for data labeling:  Oyente~\cite{rw_oyente_repo}, Mythril~\cite{rw_mythril_repo}, and Dedaub~\cite{dedaub}. 
We do not consider tools that scan smart contracts on the source code level since decompilation might result in information loss. 
The modular structure of \datatool{} makes it possible to easily extend the toolchain with other bytecode-level vulnerability detection tools \textcolor{magenta}{including \sys{}.}

\vspace{-0.4em}
\subsection{Bytecode Acquisition}
\label{sec:implementation:preclassification}
\vspace{-0.3em}
For dataset construction, \datatool{} first utilizes \textit{Infura} API to download smart contracts. We were able to download 1.155.085 bytecodes from the first 5 million Ethereum blockchain blocks. 
This number corresponds to about 96\% of the smart contracts in the targeted blocks.  
In some cases, the download resulted in an empty bytecode 0x. The possible reasons could be: (i)~The Ethereum node is not fully synced with the network, thus the bytecode is not available; (ii)~An empty contract is deployed; (iii)~The smart contract is self-destructed. 
To tackle the situation where the smart contracts' bytecode exists but cannot be downloaded via Infura API, \datatool{} uses another API provided by Dedaub tool to retrieve the missing bytecode files. 
Finally, at the end of the bytecode acquisition process, 1.156.611 smart contract bytecode files are available for \sys{}'s vulnerability analysis.  

\vspace{-1em}
\subsection{Bytecode Preprocessing}
\label{sec:implementation:preclassification}
\vspace{-0.3em}

The downloaded bytecode consists of hexadecimal digits that represent particular operation sequences and parameters. 
In the preprocessing step, \datatool{} first transforms the collected raw bytecodes to sequences of operations divided by a unique separator and removes the input parameters from the bytecode to reduce the input size.
Furthermore, it merges operations with the same functionality into one common operation. 
For instance, the similar commands \textit{PUSH1} - \textit{PUSH32} commands (represented by the bytes \textit{0x60}-\textit{0x7f}) are replaced with the \textit{PUSH} operation (represented by \textit{0x60}).
Note that some hexadecimal digits in the crawled bytecode do not correspond to any operations defined in the Ethereum Yellow Paper~\cite{ethereum-yellowpaper}. These bytes are considered as invalid operations and substituted with the value \textit{XX}.

\vspace{-0.4em}
\subsection{Labeling of Smart Contracts}
\label{sec:implementation:preclassification}
\vspace{-0.2em}

A smart contract might have multiple vulnerabilities as introduced in Section~\ref{sec:smart_contracts}. 
Each of the vulnerability detection tools used by \datatool{} is specialized for detecting a specific set of vulnerability types. 
\textcolor{magenta}{Note that besides the bytecode-level detection tools themselves, \datatool{} also store the performance metrics (e.g., F1 score) of each tool on each vulnerability type that they can detect. The performance characterizations are obtained from the previous publications~\cite{rw_oyente_repo,rw_mythril_repo,dedaub} with experts' manual inspection to ensure the correctness. 
To determine if a given smart contract has a specific vulnerability type, \datatool{} selects the detection tool that features the highest F1 score on this vulnerability among all available ones and use it for contract labeling. }

\textcolor{magenta}{\datatool{} repeats the above process for each contract and each vulnerability type.} 
We develop Python modules to perform the above task.
In the end, 1.156.611 smart contracts are labeled. It is worth repeating that since there are no dependencies among the vulnerability scanning tools, the set of vulnerability types can be easily extended with other available tools. 

\vspace{-0.2em}
\section{Implementation}
\label{sec:implementation}
\vspace{-0.2em}
In this section, we instantiate the generic design of \sys{} described in Section~\ref{sec:design} on eight
vulnerability types and elaborate on implementation details.

\vspace{-0.3em}
\subsection{Implementation of the DNN Model}
\label{sec:implementation:dl}
\vspace{-0.2em}

We begin with the instantiation of the \sys{}'s DNN model.
We utilize the \textit{tf.keras} package~\cite{tensorflow} for model building, training, and inference.
Furthermore, we devise a model serving API that takes the plain bytecode as input and returns the vulnerability detection labels as output. Our API provides end-to-end usage for the defender to update/deploy his own DNN and inspect unknown contract bytecode.
We detail the implementation procedures as follows. 

\subsubsection{Dataset Chunking}
To enable our supervised learning, a designated dataset with the preprocessed bytecode and associated labels is required. The labeled bytecode dataset for the model training process is constructed as explained in Section~\ref{sec:data}.
Furthermore, we develop a Python module to read data from the database and generate CSV files as the inputs to \sys{}'s DNN model.  
Since the size of our bytecode dataset is large, loading the full dataset into memory at one time is likely to incur out of memory (OOM) error.
To alleviate the memory constraint (Challenge (C1) in Section~\ref{sec:challenges}), we perform \textit{data chunking} as follows:
(i) The bytecode samples and the corresponding labels in the CSV file are merged into one \textit{DataFrame}; (ii)~The dataset is then shuffled and stored as a temporary file; and (iii)~We create chunks from the shuffled dataset and store the segments as chunk files based on the pre-defined chunk size (default is $1024$). The chunk files are passed directly to the DNN model.
We provide more details about our data chunking process in Appendix Section~\ref{sec:appendix:chunking}. 

Note that the collected contract data might have the class imbalance issue~\cite{rw_contractward} (Challenge (C3) in Section~\ref{sec:challenges}). In our work, we construct a balanced training set to ensure that each batch of data fed into \sys{}'s DNN model has a comparable number of vulnerable and safe contracts. Details about our dataset balancing is given in Section~\ref{sec:evaluation:dataset}. 

\subsubsection{Model Building and Training}
To realize \sys{}'s DNN-based vulnerability scanning approach, we instantiate a Multi-Output-Layer (MOL) DNN model based on our generic architecture explained in Section~\ref{sec:approach:deep-learning}.

%% actual model 
%\vspace{-1.5em}
\begin{figure}[ht!]
  \centering
  \includegraphics[width=0.86\columnwidth]{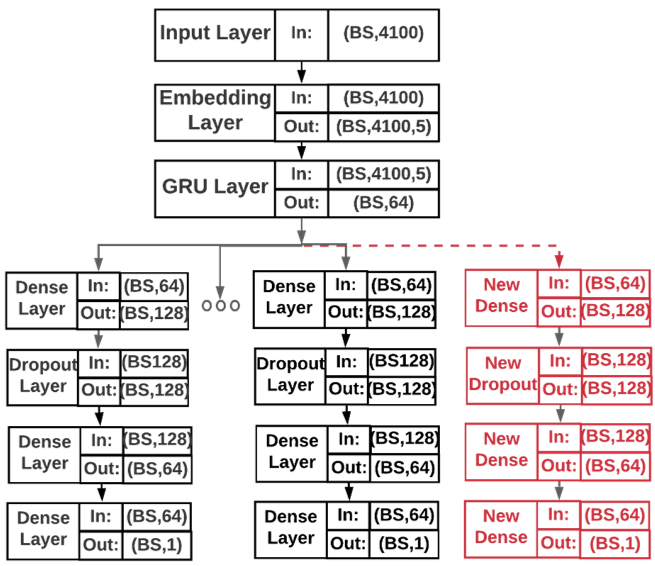}
  \vspace{-0.8em}
  \caption{Multi-output DNN architecture of \sys{} for concurrent detection of multiple vulnerability types. Here, $BS$ is the batch size, the rightmost red branch denotes adding a new branch for transfer learning on new vulnerabilities. }
    \label{fig:implementation:mol-gru}
    \vspace{0.8em}
\end{figure}

Figure~\ref{fig:implementation:mol-gru} visualizes the actual model implemented in our experiments. We first build a MOL-DNN model with six output branches for main model training and then extend it with two new branches for transfer learning.
Recall that our objective is to enable concurrent detection of multiple vulnerability types. To this end, we construct a recurrent neural network by stacking embedding layer, GRU layers, Dropout layers, and Dense layers to process sequential bytecode inputs. 
The feature extractor is concatenated with multiple branches to output the probability of having each specific vulnerability.  
After building the MOL-DNN model, we specify the model hyper-parameters as shown in Table~\ref{tab:model_param}, and launch the training process as depicted in Figure~\ref{fig:global}.
Since the bytecode dataset is chunked to avoid OOM errors, \sys{}'s DNN model iterates over all chunks multiple times during model training.

%\vspace{-1em}
\begin{table}[ht!]
\centering
\scalebox{0.96}{
\begin{tabular}{|l|l|}
\hline
\thead{\textbf{Variable}} & \thead{\textbf{Setting}} \\ \hline
Layer Type & Embedding, GRU, Dense, Dropout \\ \hline
\#Hidden Unites& GRU:64, Dense:[128,64,1]  \\ \hline
Optimizer &  Adam \\ \hline
Loss Function & Binary Cross-Entropy \\ \hline
Learning Rate & 0.001 \\ \hline
Dropout Values  &  0.2 \\ \hline
\#Local\_Epochs & 1  \\ \hline
\#Global\_Epochs & 1  \\ \hline
Batch Size     & 32 \\ \hline
MAX Seq. Length & 4100 \\ \hline
\end{tabular}
}
\vspace{1.6em}
\caption{Model Hyper-parameters.}
\label{tab:model_param}
\vspace{-1.6em}
\end{table}

\begin{figure}[ht!]
  \centering
  \includegraphics[width=0.77\columnwidth]{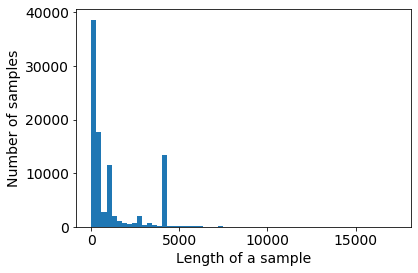}
  \vspace{-1.5em}
  \caption{Distribution of bytecodes length in the dataset.}
  \label{fig:dataset-length-distribution}
  \vspace{0.8em}
\end{figure}

We define `\#Global\_Epochs' to be the total number of times that the full training set is iterated by our DNN model and `\#Local\_Epochs' as the number of times that each chunk is used to update the model.
Before the chunked data are passed to the model's input layer, the bytecode sequence needs to be vectorized.
This is realized by a \textit{tokenizer}, which transforms the hexadecimal data into numeric vectors. 
After tokenization, a hyper-parameter \textit{MAX\_SE\-QUENCE\-\_LENGTH} is applied to the input vectors. Sequences shorter than this length are zero-padded, while sequences longer than this length are truncated. 
We empirically study the distribution of the bytecode length and show the results in Figure~\ref{fig:dataset-length-distribution}. The hyperp-arameter \textit{MAX\_SE\-QUENCE\-\_LENGTH} is set to 4100 to ensure more than $98.5\%$ of the contracts are not truncated\footnote{Both excessive padding and truncation may worsen performance. For instance, truncated contracts may end up being mislabeled as benign if a vulnerability resides in the truncated part of the contract.}.

%\vspace{-1.1em}

The tokenized data are then passed to the DNN model for training. We assess the classification accuracy of the model after each Local Epoch (L\_Ep., train on one chunk) and each Global Epoch (G\_Ep., train on all chunks). The evaluation results are stored in a metric history object to keep track of the progress.  
In the testing phase, the remaining (unseen) data chunks are passed to the model to compute the detection metrics. 
At the end of the model training stage, we save the converged MOL-DNN model, the used tokenizer, and the evaluation metrics files to wrap the model as an API service.

\vspace{-0.3em}
\subsection{Model Serving API}
\label{sec:implementation:deeplearning:modelapi}
\vspace{-0.2em}
Our trained DNN model can detect pre-defined vulnerability types in smart contracts. We wrap the model within an API to ensure that we can serve predictions to end-users on the fly. We utilize Flask~\cite{flask} and devise a Python module to provide a REST API endpoint for running model inference on bytecode files.  
Our API also performs automated bytecode transformation (input preprocessing) to remove the manual efforts for the defender. 
The learned \sys{} model and associated configurations are passed to our API. 
The Python module creates two different API endpoints. The first endpoint shows the configuration passed to the module and the second one triggers model inference. 
Listing~\ref{lst:model-api-request-body} shows an example of the plain bytecode of the smart contracts, which is passed to the second endpoint for vulnerability detection.

\vspace{-0.4em}
\begin{lstlisting}[caption={Sample request body when calling \sys{}'s prediction endpoint.},language=json,label={lst:model-api-request-body},firstnumber=1,frame=tblr
,basicstyle=\fontsize{8}{9}\selectfont
]
{"smart_contract": "606060405236150100000..."}
\end{lstlisting}

In the second endpoint, the bytecode of the input smart contract is vectorized using the same tokenizer as the one used in \sys{}'s model training step. The processed sequence is then fed as the input to the trained MOL-DNN model for vulnerability detection. In addition to the detected vulnerability types, the prediction time of \sys{} is tracked and shown to the user. An example response from \sys{}'s API endpoint is shown in Listing~\ref{lst:model-api-prediction}.

\vspace{-0.4em}
% \begin{lstlisting}[caption={Sample request response containing the model prediction when calling the prediction endpoint.},label={lst:model-api-prediction},language=json,firstnumber=1,frame=tblr, basicstyle=\fontsize{8}{9}\selectfont]
\begin{lstlisting}[caption={\sys{}'s response to the sample request when calling our prediction endpoint. The response includes the analysis results of multiple vulnerability types. },label={lst:model-api-prediction},language=json,firstnumber=1,frame=tblr, basicstyle=\fontsize{8}{9}\selectfont]
{"prediction": {
    "ASSERT_VIOLATION": 0.0001,
    "ACCESSIBLE_SELFDESTRUCT": 0.9998,
    "DoS (UNBOUNDED_OP)": 0.9996,
    "MULTIPLE_SENDS": 0.0012,
    "TAINTED_SELFDESTRUCT": 0.9998,
    "CALLSTACK": 0.9995,
    "MONEY_CONCURRENCY": 0.0013,
    "REENTRANCY": 0.0009}, 
 "prediction_time in_second": "0.02"}
\end{lstlisting}
\vspace{-0.3em}
\section{Evaluation}
\label{sec:evaluation}
% \vspace{-0.3em}
We assess \sys{} on the large-scale smart contract dataset built as described in Section~\ref{sec:data}.
In this section, we explain our experimental setup and the evaluation metrics to characterize the performance of \sys{}'s DNN model.  

\vspace{-0.3em}
\subsection{Dataset}
\label{sec:evaluation:dataset}
% \vspace{-0.2em}
To build our dataset, we collected $\sim1.2$ million smart contracts from the first 5 million Ethereum blockchain blocks using \datatool{}, as explained in Section~\ref{sec:implementation:preclassification} and label them accordingly. Note that vulnerability scanners can normally detect multiple vulnerability types, thus there might also be overlap in coverage. 
For instance, a reentrancy bug is detected by all three tools, Oyente, Mythril, and Dedaub.

To select only one label for redundantly labeled vulnerability types, we first compute respective F1 scores of different tools~\cite{metric_web} based on the true positives, false positives, and false negatives reported in the papers.
The tool with the highest F1 score on this particular vulnerability class is then used to determine if the collected smart contracts are vulnerable to this vulnerability class.
For instance, Oyente~\cite{rw_oyente_repo} is used to detect the reentrancy bug since it yields the highest score. 

For the purpose of evaluation, we include eight vulnerability types in our dataset. 
However, as discussed in Section~\ref{sec:transfer}, we stress that our approach is not limited only to eight classes and can be easily extended with new attacks. 

In the following, we define eight vulnerability types included in our evaluation and refer to vulnerability categories presented in Section~\ref{sec:smart_contracts} for more detailed vulnerability description. 

\begin{itemize}
%\vspace{-0.8em}
\item \textbf{Callstack Depth [cl. 1]}: 
This vulnerability class belongs to the category \textbf{Programming Error} (cf. Section~\ref{sec:smart_contracts}) and exploits the stack size limit issues of the EVM.

\item \textbf{Reentrancy [cl. 2]}: % detected by Oyente:
% As explained in Section~\ref{}. https://swcregistry.io/docs/SWC-107
Reentrancy bugs are caused by \textbf{External Calls} and allow an attacker to drain funds, as explained in Section~\ref{sec:smart_contracts}.

\item \textbf{Multiple Sends [cl. 3]}:
This class hinges on the exploitation of a smart contract's \textbf{Execution Costs} to induce DoS.

\item \textbf{Accessible selfdestruct [cl. 4]}: 
This \textbf{Programming Error} can be exploited to terminate a contract such that the remaining funds are sent to a predefined address.

\item \textbf{DoS (Unbounded Operation) [cl. 5]}: 
An attacker can exploit the limited \textbf{Execution Costs} of a smart contract when the execution time is dependent on input from an external caller.

\item \textbf{Tainted selfdestruct [cl. 6]}: 
This vulnerability is an extension of vulnerability class 4 (cl. 4). The attacker here can set to which address the remaining balance of the smart contract is sent.

\item \textbf{Money concurrency [cl. 7]}:
This vulnerability is also known as Transaction Ordering Dependence (TOD) and belongs to the \textbf{Influence by Miners} category.

\item \textbf{Assert violation [cl. 8]}:
This \textbf{Programming Error} leads to a constant error state of the smart contract, which can be exploited by an attacker.
\end{itemize}

\begin{figure}[ht!]
\centering
\includegraphics[width=0.88\columnwidth]{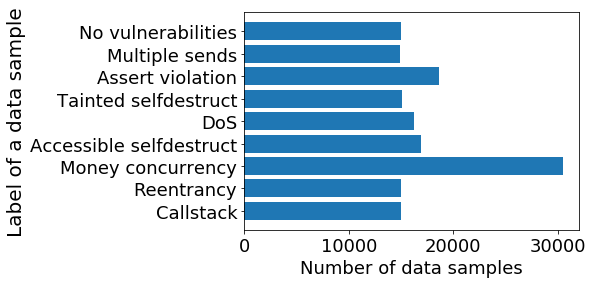}
\vspace{-1em}
\caption{Vulnerability class distribution of our dataset. 
}
\label{fig:dataset-distribution}
% \vspace{-0.7em}
\vspace{0.7em}
\end{figure}

For each vulnerability class, we select at least 15.000 samples with this specific vulnerability from our raw dataset ($\sim 1.2$ million contracts) and concatenate them to construct the vulnerable contract set with the equally-sized distribution. 
We empirically set the minimal sample number to 15.000 since this is the smallest size of the well-represented vulnerability types in our dataset. The above dataset was then completed with 15.000 completely clean smart contracts where no vulnerabilities were detected by the tools used in this work (Section~\ref{sec:implementation:preclassification}).
Figure~\ref{fig:dataset-distribution} shows the vulnerability class distribution of our balanced, labeled dataset.  
One can see from Figure~\ref{fig:dataset-distribution} that our dataset construction method solves the challenge (C3) formulated in Section~\ref{sec:challenges}.

%% vuln class distirbution 
%\vspace{-0.6em}
    
It is worth mentioning that the actual size of our labeled dataset is 93.497 samples instead of $15000 \times (8+1)$. This is because \sys{} formulates vulnerability detection as \textit{multi-label classification}, meaning that a contract might have multiple labels and repeatedly appear in the selection of several vulnerability types described above.
We take $80\%$ and $20\%$ of the labeled dataset to construct the training and test dataset for \sys{}. Note that we set aside $10\%$ of the training data as the validation data to prevent model over-fitting.

% \vspace{-0.4em}
\subsection{Evaluation Metrics}
\label{sec:evaluation:metrics}
\vspace{-0.2em}
\sys{} provides concurrent detection of multiple vulnerability types. We evaluate the performance of \sys{}'s multi-output DNN model with F1 score, precision, and recall. 
In addition, we calculate two of the most common performance metrics, hamming loss, and Jaccard similarity~\cite{book:MML}.  
We detail each metric as follows.

\subsubsection{Base Values}
The results of true positives (TP), true negatives (TN), false positives (FP), and false negatives (FN) are the base values to compute other metrics. 
The true values represent the number of correctly predicted results, which can be either true positive or true negative. 
The false values indicate that the DL model gives the wrong outputs~\cite{deep-learning-glossary}.

\vspace{0.1em}
\subsubsection{Precision and Recall}
The precision metric describes the ratio of truly positive values to all positive predictions. 

This indicates the reliability of the classifier's positive prediction~\cite{deep-learning-metrics}. 
The recall (or sensitivity) metric shows the proportion of actual positives that are correctly classified.  
The formulas to compute these two metrics are given below:

 \resizebox{.96\linewidth}{!}{
   \begin{minipage}{\linewidth}
   \begin{align}
   \label{eq:precision_recall}
     Precision = \frac{TP}{TP + FP},~Recall = \frac{TP}{TP + FN}. 
   \end{align}
   \end{minipage}
 }

\vspace{0.1em}
\subsubsection{F1 Score}
The F1 score metric is commonly used in information retrieval and it quantifies the overall decision accuracy using precision and recall.  
The F1 score is defined as the \textit{harmonic mean} of the precision and recall:

\resizebox{.96\linewidth}{!}{
   \begin{minipage}{\linewidth}
   \begin{align}
   \label{eq:f1_score}
      F1\_score = \frac{2 * Precision * Recall}{Precision + Recall}. 
   \end{align}
  \vspace{0.03em}
   \end{minipage}
}
The best and the worst value of the F1 score is 1 and 0, respectively. The F1 score can be calculated for each class label or globally~\cite{sklearn}. In our evaluation, we use the weighted F1 score where the per-class F1 scores are weighted by the number of samples from that class~\cite{metric_web}.
 
\subsubsection{Jaccard Similarity} In multi-label classification, Jaccard similarity (Jaccard index) is defined as the size of the intersection divided by the size of the union of two label sets. It is used to compare the set of predicted labels for a sample to the corresponding set of true labels. It ranges from 0 to 1 where 1 is the perfect score. The Jaccard similarity does not consider the correct classification of negatives~\cite{deep-learning-metrics} and can be computed as follows: \\

 \resizebox{.96\linewidth}{!}{
   \begin{minipage}{\linewidth}
   \begin{align}
   \label{eq:jaccard}
      Jaccard\_Similarity = \frac{TP}{TP + FP + FN}.
   \end{align}
 %   \vspace{0.05em}
   \end{minipage}
 }

\vspace{0.1em}
\subsubsection{Hamming Loss}
The Hamming loss gives the percentage of wrong labels to the total number of labels. A lower hamming loss indicates the better performance of a model. For an ideal classifier, the hamming loss is 0. In multi-label classification, the hamming loss is defined as the \textit{hamming distance}~\cite{norouzi2012hamming} between the ground-truth label $y$ and the prediction value $\hat{y}$: \\ 
% {\footnotesize
%  \begin{equation}
%      \mathcal{L}_{Hamming} = \frac{\#Mismatch(y,~\hat{y})}{Length(y)}.
%  \end{equation}
%  }
%\begin{comment}
\resizebox{.96\linewidth}{!}{
  \begin{minipage}{\linewidth}
  \begin{align}
  \label{eq:hamming}
     \mathcal{L}_{Hamming} = \frac{\#Mismatch(y,~\hat{y})}{Length(y)}.
  \end{align}
  \vspace{0.05em}
  \end{minipage}
}
%\end{comment}

\noindent where $Length(y)$ is the total number of vulnerability types.

\vspace{0.1em}
\subsubsection{DNN Loss Function}
The loss function is a crucial part of DNN training since the training process aims to minimize the loss for obtaining a high task accuracy. 
As such, the loss value quantifies how well a classifier performs on the given dataset. In our experiments, we use \textit{Binary Cross-Entropy} (BCE) loss to train \sys{}'s multi-output DNN model. Given the expected value $y$ and the prediction $\hat{y}$, the BCE loss is computed as:  \\
% {\footnotesize
%  \begin{equation}
%   \mathcal{L}_{BCE}(y, \hat{y}) = -(y log(\hat{y}) + (1 - y) log (1 - \hat{y})). 
%  \end{equation}
%  }
% \vspace{-2em}
 \resizebox{.96\linewidth}{!}{
   \begin{minipage}{\linewidth}
   \begin{align}
   \label{eq:bce_loss}
      \mathcal{L}_{BCE}(y, \hat{y}) = -(y log(\hat{y}) + (1 - y) log (1 - \hat{y})). 
   \end{align}
% %   \vspace{0.05em}
   \end{minipage}
 }

% \paragraph{Subset Accuracy}
% The subset accuracy calculates the accuracy of the whole output in the case of multi-label models~\cite{tensorflow}. Therefore, it is defined as:

% \begin{equation}
%   Subset\_Acc = \frac{subset\_true\_pred}{subset\_true\_pred + subset\_false\_pred}
% \end{equation}

% \vspace{-0.3em}
\subsection{Experimental Setup}
\label{sec:evaluation:setup}
% \vspace{-0.2em}
All of our experiments are conducted on a machine with Arch Linux OS having AMD Ryzen 3 3200G and NVIDIA GeForce RTX 3090 GPU with 32 and 24 GB of RAM, respectively. The software versions are as follows: Tensorflow 2.3.1, CUDA 11.1, NVIDIA driver 455.38, cuDNN 8.0.5.39-1, and kernel 5.4.77-1.

\vspace{-0.3em}
\subsection{Evaluation Results}
\label{sec:evaluation:results}
% \vspace{-0.2em}
%%
\subsubsection{Classifier Learning}
\begin{figure}[h]
  \centering
    \subfloat[\sys{}'s BCE loss on training and validation set.]{
     \label{figur:metrics-training:loss}
     \includegraphics[width=0.68\columnwidth]{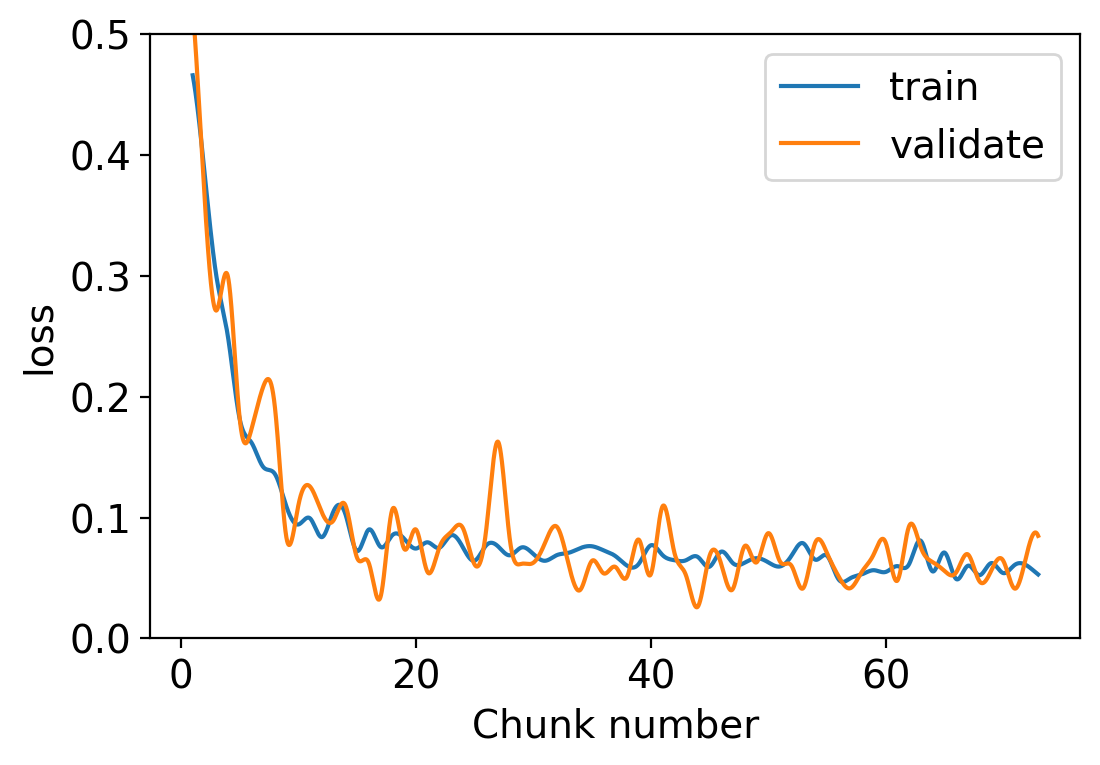}
  }
  
%   \hfill
  \vspace{-0.8em}
  \subfloat[\sys{}'s F1 score on training and validation set.]{
     \label{figur:metrics-training:f1}
     \includegraphics[width=0.68\columnwidth]{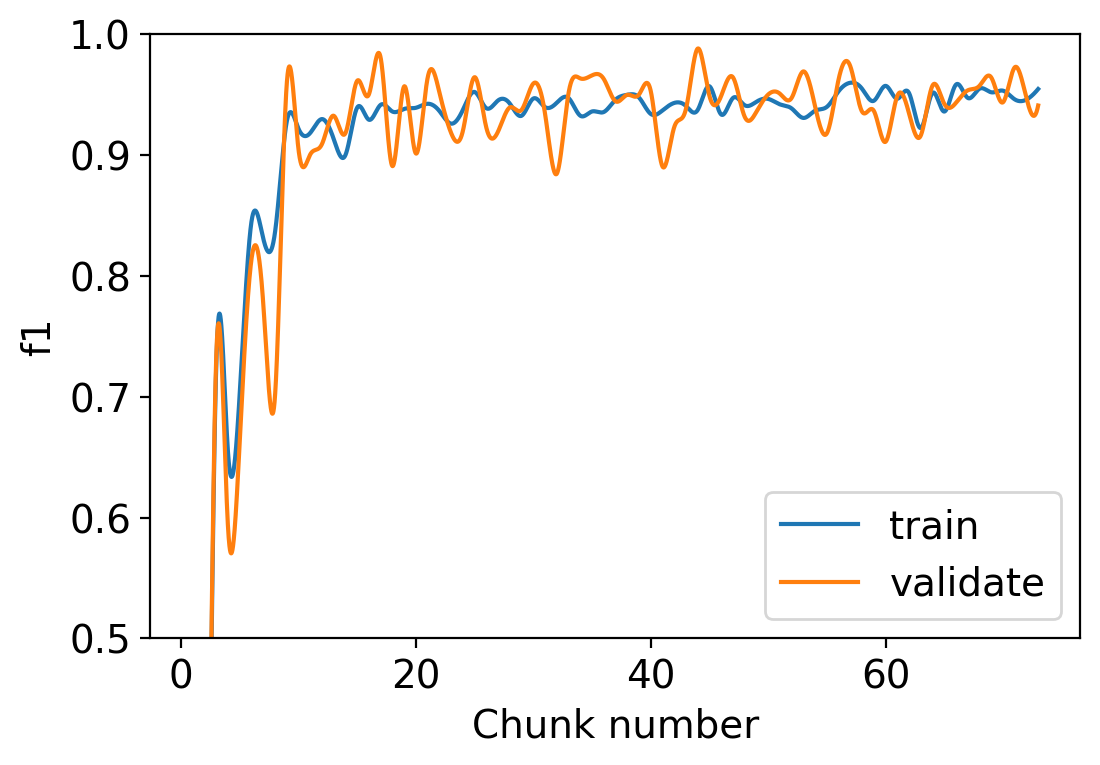}
  }
  \vspace{-0.5em}
  \caption{The model learning process of \sys{}. 
  }
  \label{fig:metrics-training-part1}
\end{figure}

% \textbf{Classifier Learning.} 
We train \sys{}'s model on the training set with hyper-parameters listed in Table~\ref{tab:model_param} and assess it on the test set {(consisting of $18.700$ contracts)}.
To corroborate the detection effectiveness of \sys{}, we plot the learning curves of our multi-output DNN 
in Figure~\ref{fig:metrics-training-part1}. 
The training and the validation curves demonstrate the \textit{time-evolving} performance of \sys{} for supervised learning and the \textit{generalization} capability of the model, respectively. 
The learning curves show that our DNN model can achieve an average F1 score higher than $95\%$ on both the training and validation set.  

%% ===== learning curve ===== %% 
% \vspace{-1.5em}

\vspace{0.8em}
\subsubsection{Sensitivity to Training Configurations}  
\sys{}'s training pipeline is described in Section~\ref{sec:implementation:dl}.
We use the MOL-DNN architecture in Figure~\ref{fig:implementation:mol-gru} with six vulnerability branches in \sys{}'s training phase. 
This model has a total number of $115,846$ learnable parameters.
We investigate whether the detection performance of \sys{} can be improved by increasing the number of global and local epochs.
To characterize \sys{}'s performance, we compute the metrics defined in Section~\ref{sec:evaluation:metrics} and summarize the results in Table~\ref{tab:differentiterationepoch}. 
It can be seen that \sys{}'s accuracy has a slight increase with a longer training time. Since the accuracy improvement is minor w.r.t. the large increase of training time, we use the first configuration (G\_{Ep}=1, L\_{Ep}=1) as our default setting for the rest of our evaluations.

%\vspace{-0.3em}
%% ==== perf. vs. training configs === %%
\begin{table}[ht!]
	\centering
	\scalebox{0.95}{
\begin{tabular}{|l|l|l|l|l|l|}
\hline
\multicolumn{2}{|l|}{\thead{Metrics (Avg.)}}               & \thead{G\_Ep.:1 \\ L\_Ep.:1} & \thead{G\_Ep.:1 \\ L\_Ep.:3} & \thead{G\_Ep.:3 \\ L\_Ep.:1} & \thead{G\_Ep.:3 \\ L\_Ep.:3} \\ \hline
\multirow{6}{*}{\rotatebox[origin=c]{90}{Training}} & Loss &            0.06              &         0.05                 &          0.06                &     0.05     \\ \cline{2-6} 
                          & Precision                      &            0.99              &         1.00                 &          0.99                &     0.97     \\ \cline{2-6} 
                          & Recall                         &            0.91              &         0.90                 &          0.90                &     0.95     \\ \cline{2-6} 
                          & F1 score                       &            0.95              &         0.95                 &          0.94                &     0.96     \\ \cline{2-6} 
                          & Hamming Loss                   &            0.02              &         0.02                 &          0.02                &     0.01     \\ \cline{2-6} 
                          & Jaccard Similarity             &            0.90              &         0.90                 &          0.89                &     0.92     \\ \hline
\multirow{6}{*}{\rotatebox[origin=c]{90}{Test}}     & Loss &            0.06              &         0.05                 &          0.05                &     0.05     \\ \cline{2-6} 
                          & Precision                      &            0.98              &         0.99                 &          0.99                &     0.97     \\ \cline{2-6} 
                          & Recall                         &            0.90              &         0.91                 &          0.91                &     0.95     \\ \cline{2-6} 
                          & F1 score                       &            0.95              &         0.95                 &          0.95                &     0.96     \\ \cline{2-6} 
                          & Hamming Loss                   &            0.02              &         0.02                 &          0.02                &     0.01     \\ \cline{2-6} 
                          & Jaccard Similarity             &            0.90              &         0.90                 &          0.90                &     0.92     \\ \hline
\multirow{2}{*}{\rotatebox[origin=c]{90}{Time}} & Training (H:M)   &   5:44              &        17:11                 &         17:40                &    49:41     \\ \cline{2-6} 
                          & Prediction (Sec.)               &            0.02              &         0.02                 &          0.02                &     0.02     \\ \hline
\end{tabular}
}
\vspace{2em}
\caption{The averaged performance of our multi-label classification model across initial vulnerability types (cl. 1-6) with different G\_Ep. and L\_Ep. configurations.
% Global Epoch (G\_Ep.) and Local Epoch (L\_Ep.) configuration. 
}
\label{tab:differentiterationepoch}
\end{table}

%%
%\vspace{0.8em}
\subsubsection{Class-wise Detection Performance} 
We describe \sys{}'s overall performance on the vulnerability types introduced in Section~\ref{sec:evaluation:dataset}. 
Here, we provide a fine-grained insight into \sys{}'s capability on each vulnerability class. 
Table~\ref{tab:class-metrics} shows the class-specific metrics obtained by our MOL-DNN model. 
\sys{} achieves an average of $97\%$ precision and $95\%$ F1 score across all six vulnerability types. 
{In the case of the Accessible selfdestruct vulnerability (cl. 4), \sys{}'s recall score is relatively lower than others. The low recall value indicates that \sys{} yields more false negatives on this vulnerability class according to Equation~(\ref{eq:precision_recall})}.

%% === class-specific perf. === %%
\begin{table}[ht!]
	\centering
	\scalebox{0.95}{
\begin{tabular}{|l|l|l|l|l|l|l|l|l|}
\hline
\multirow{2}{*}{\thead{Metrics}} & \multicolumn{6}{l|}{ \thead{Initial vulnerability types}} & \multicolumn{2}{l|}{\begin{tabular}[c]{@{}l@{}} \thead{New Vuln.\\ Classes}\end{tabular}} \\ \cline{2-9} 
                    & cl. 1     & cl. 2     & cl. 3     & cl. 4   & cl. 5   &  cl. 6   &cl. 7  & cl. 8  \\ \hline
Loss                &   0.08    &   0.05    &   0.07    &  0.09   &  0.01   &   0.09   & 0.14  &  0.08  \\ \hline
Precision           &   0.99    &   0.97    &   0.99    &  0.99   &  1.00   &   0.98   & 0.95  &  0.98  \\ \hline
Recall              &   0.88    &   0.96    &   0.91    &  0.85   &  0.99   &   0.88   & 0.90  &  0.90  \\ \hline
%Jaccard             &   0.87    &   0.92    &   0.90    &  0.84   &  0.99   &   0.87   & 0.86  &  0.88  \\ \hline
F1 score            &   0.93    &   0.96    &   0.95    &  0.91   &  0.99   &   0.93   & 0.92  &  0.93  \\ \hline
FPR                 &   0.00    &   0.01    &   0.00    &  0.00   &  0.00   &   0.00   & 0.02  &  0.00  \\ \hline
FNR                 &   0.12    &   0.06    &   0.10    &  0.15   &  0.01   &   0.12   & 0.10  &  0.10  \\ \hline
\end{tabular}
}
\vspace{2em}
   \caption{Class-specific metrics (for unseen/test data) retrieved by our multi-output model for initial training phase (cl. 1-6) and after transfer learning phase (cl. 7-8). %\textcolor{red}{need to update FNR.}
   }   \label{tab:class-metrics}
\end{table}

%\vspace{0.3em}

\subsection{Extensibility/Generalizability Performance}

To corroborate the extensibility of \sys{}, we first train a MOL-DNN model on six vulnerability types (cl. 1-6), and then perform transfer learning on the remaining two vulnerability types (cl. 7-8). The details of the transfer learning procedures can be found in Section~\ref{sec:transfer}. 
The two newly added branches (for cl. 7-8) have in total $33,282$ trainable parameters. For transfer learning, we used the same hyper-parameters listed in Table~\ref{tab:model_param} and trained the model on vulnerability types 7 and 8 simultaneously.
This process takes only $2$ hours and $43$ minutes, which is $47.39\%$ of the time used by model training with six vulnerabilities. 
The detection result on the new vulnerability types are shown in the last two columns of Table~\ref{tab:class-metrics}.
It can be seen that \sys{} achieves $92\%$ and $93\%$ F1 score on the two new vulnerabilities cl. 7 and cl. 8, respectively. 
The empirical results corroborate that \sys{} is extensible to new attacks by enabling lightweight and effective transfer learning.

\textcolor{magenta}{We also demonstrate the superior efficiency and effectiveness of \sys{}'s innovative multi-output architecture compared to the existing ML-based detection techniques. 
In this comparison experiment, we define a regular RNN consisting of the three top layers and two branches shown in Figure~\ref{fig:implementation:mol-gru} as the baseline model. This new model is trained \textit{from scratch} on the two new vulnerability types. For \sys{}, we use the pre-trained feature extractor obtained from the main training stage and append two new branches of layers to it. As we explain in Section~\ref{sec:transfer}, only the parameters in the new vulnerability branches are trained during transfer learning.}

\textcolor{magenta}{Note that the size of the transfer learning dataset is considerably smaller than the data size in the main training phase since the new smart contracts are collected in a shorted time period. The limited training data raises concerns about model \textit{under-fitting} if the amount of trainable parameters is too large. We show the empirical comparison results in Table~\ref{tab:tl_comparison} when adapting the ML model to the new vulnerability types (cl. 7 and cl. 8). It can be seen that \sys{} reduces the training time by $\sim 43\%$ while yielding a similar F1 score compared to the baseline. The advantage of \sys{} is derived from our multi-output RNN design which decomposed the contract vulnerability problem into two sub-tasks: contract representation learning (performed by feature extractor), and vulnerability pattern identification (performed by each branch). As such, when extending the ML-based detector to new vulnerability types, \sys{} can focus on distinguishing the new vulnerability pattern without \textit{`re-learning'} the contract representation.}

\vspace{-0.3em}
\begin{table}[ht!]
\centering
\scalebox{0.96}{
\begin{tabular}{|c|c|c|c|c|}
\hline
\textbf{\begin{tabular}[c]{@{}c@{}}Transfer \\ Learning\end{tabular}} & \textbf{FPR} & \textbf{FNR} & \textbf{F1 score} &  \textbf{\begin{tabular}[c]{@{}c@{}}Training Time \\ (H:M)\end{tabular}} \\ \hline
ESCORT & 0.02 & 0.10 & 0.93 & 2:43 \\ \hline
Baseline & 0.01 & 0.10 & 0.94 & 4:46 \\ \hline
\end{tabular}
}
\vspace{1.8em}
\caption{Performance comparison of transfer learning between \sys{} and baseline (training-from-scratch). \label{tab:tl_comparison}}
\end{table}

\vspace{-2.3em}
\section{Related Work} 
\label{sec:related}
% \vspace{-0.3em}
Various vulnerability inspection tools have been developed for smart contracts to ensure the security of the cryptocurrency system. We categorize the existing detection techniques based on their working mechanism and discuss each type below. 

%% 
% \vspace{-0.3em}
\subsection{Static Vulnerability Detection Methods}
\label{sec:related:static}
% \vspace{-0.3em}
Static detection techniques analyze the smart contract in a static environment by examining its source code or bytecode. 

\subsubsection{Information Flow Analysis-based}
{Slither}~\cite{feist2019slither} uses \textit{taint analysis}~\cite{tripp2009taj} to detect vulnerabilities in Solidity source code. 
It can find nearly all vulnerabilities related to the user inputs or critical data flows while the inspection time might be prohibitively long.
{Dedaub's Contract-Library} by Dedaub~\cite{rw_dedaub_contract_library_page} provides multiple different features via an online API. It collects bytecode of the smart contracts and performs vulnerability classifications using tool MadMax~\cite{rw_dedaub_mad_max_paper} that performs flow and loop analysis to detect gas-focused vulnerabilities~\cite{brent2020ethainter}. 

\subsubsection{Symbolic Execution-based}
{Oyente}~\cite{ethereum-background-03} detects vulnerabilities in the source code or bytecode of Solidity contracts using \textit{symbolic execution}. 
Symbolic execution represents the program's behavior as built formula and uses symbolic inputs to decide if a certain path can be reached~\cite{king1976symbolic}.
As such, its performance depends on the number of explored paths and the program's complexity~\cite{rw_tech_01, rw_tech_02}. 
Oyente constructs the control flow graph of the contract and uses it to create inputs for symbolic execution. 
{Manticore}~\cite{rw_manticore_paper} analyzes the contract by repeatedly executing symbolic transactions against the bytecode, tracking the discovered states, and verifying code invariants~\cite{rw_manticore_repo}.
{Securify}~\cite{rw_securify_paper}
first obtains the contract's semantic information by performing symbolic analysis of the dependency graph, then checks the predefined compliance and violation patterns for vulnerability detection. 
teEther~\cite{teether} searches for certain critical paths in the control flow graph of the smart contract and uses symbolic execution for vulnerability identification.

\subsubsection{Logic Rules-based}
Vandal~\cite{brent2018vandal} is a logic-driven static program analysis framework.  
It converts the low-level EVM bytecode to semantic logic relations and describes the security analysis problems with logic rules. The datalog engine executes the specifications for input relations and outputs the vulnerabilities. % and locations in bytecode. 
{eThor}~\cite{ethor} is a static analysis technique built on top of reachability analysis achieved by Horn clause resolution.  
{NeuCheck}~\cite{NeuCheck} adopts a syntax tree in a syntactical analyzer to transform source code of smart contracts to an intermediate representation (IR). Vulnerabilities are identified by searching for detection patterns in the syntax tree. 
{SmartCheck}~\cite{smartcheck} converts the Solidity source code to XML-based IR and verifies it against detection patterns defined in XPath language~\cite{xpath}. 

\subsubsection{Composite Methods-based}
Mythril~\cite{rw_mythx_article} combines multiple vulnerability detection approaches, including symbolic execution, taint analysis, and Satisfiability Modulo Theories (SMT). SMT solving converts the contract to SMT constraints to reveal program flaws. 
{Zeus}~\cite{zeus} uses symbolic model checking, abstract interpretation, and constrained horn clauses to verify contracts' security. 
{Osiris}~\cite{osiris} combines symbolic execution and taint analysis to precisely identify integer bugs in smart contracts. 

\subsection{Dynamic Vulnerability Detection Methods}
\label{sec:related:dynamic}
% \vspace{-0.3em}
Dynamic testing techniques execute the program and observe its behaviors to determine the vulnerability's existence. 

\subsubsection{Fuzzing-based} {MythX}~\cite{rw_mythx_page} 
combines synthetic execution and \textit{code fuzzing}. It provides a cloud-based API for developers to inspect smart contracts.
Fuzzing~\cite{godefroid2012sage} is a testing method that attempts to expose the vulnerabilities by executing the program with invalid, unexpected, or random inputs. The brute-force nature determines that fuzzing incurs large runtime overhead and might have poor code coverage due to its dependency on the inputs~\cite{rw_tech_02}.
{ReGuard}~\cite{ethereum-background-05} is another fuzzing tool specialized in the Reentrancy bug. It creates an IR for the smart contract. A fuzzing engine is used to generate random byte inputs and analyze the execution traces for reentrancy bugs detection.  
{ContractFuzzer}~\cite{contractfuzzer} generates fuzzing inputs based on the ABI specifications of smart contracts. Test oracles are defined to monitor and analyze the contract's runtime behaviors for vulnerability detection. 
{Echidna}~\cite{echidna} is a fuzzer that generates random tests to detect violations in assertions and custom properties. 
{ILF}~\cite{ilf} uses symbolic execution to generate contract inputs and employs imitation learning to design a neural network-based fuzzer from symbolic execution.
{sFuzz}~\cite{nguyen2020sfuzz} is an adaptive fuzzer for smart contracts that combines the AFL fuzzer and multi-objective strategy to explore hard-to-cover branches. 
{Harvey}~\cite{wustholz2020harvey} is a greybox fuzzer that predicts new inputs to cover new paths and fuzzes the transaction sequence in a demand-driven manner.

\subsubsection{Validation-based}  ContractLarva~\cite{rw_contractlarva_paper} is a runtime verification tool for smart contracts where a violation of defined properties can lead to various handling strategies, such as a system stop. These properties can include undesired event traces of control or data flow. 
Maian~\cite{rw_maian_paper,rw_maian_repo} combines symbolic analysis and concrete validation to inspect the smart contract's bytecode. In concrete validation, the contract is executed on a fork of Ethereum for tracing and validation. By passing symbolic inputs to the contract, the execution trace is analyzed to identify the vulnerabilities.
Sereum~\cite{rodler2018sereum} uses runtime monitoring and verification to protect existing smart contracts against reentrancy attacks without modifications or semantic knowledge of the contracts. 
It detects inconsistent states in the contract via dynamic taint tracking and data flow monitoring during contract execution. %, thus preventing reentrancy attacks. 

\vspace{-1.2em}
\subsection{Machine Learning for Vulnerability Detection}
% \vspace{-0.3em}
Several works have attempted to perform automated contract scanning using machine learning techniques. We discuss their working mechanisms and limitations below.

\vspace{0.2em}
% {\tikz\draw[black,fill=black] (-0.5em,-0.5em) rectangle (-0.2em,-0.2em);}
\textbf{ContractWard.}
ContractWard detects smart contracts vulnerability in the \textit{opcode-level} by extracting \textit{bigram} features from the simplified opcode and training individual binary ML classifiers for each vulnerability class~\cite{rw_contractward}. The paper targets six vulnerabilities and experiments with Random Forests, K-Nearest Neighbors, SVM, AdaBoost, and XGBoost classifiers. 

Compare to our work, ContractWard has three main limitations:
\textit{(i) Requires source code of smart contracts}. This approach analyzes smart contracts with opcodes. To do so, it decompiles source codes and converts them to opcodes. It is worth mentioning that decompilation might result in information loss. \textit{(ii) Not extensible to new exploitation attacks}. This paper uses `One vs. Rest' algorithms and designs separate ML models to detect each vulnerability class. This means that supporting a new vulnerability class requires training a new ML model from scratch, which is costly.  
\textit{(iii) Not scalable to long contracts}. The \textit{bigram} language model has a short window size in the Markov chain model. This determines that ContractWard is not scalable to long contracts and cannot capture long-term dependency in the code. However, using an n-gram model with a larger window size increases the feature size, thus complicating model training due to the high dimensionality of data.

\vspace{0.2em}
% {\tikz\draw[black,fill=black] (-0.5em,-0.5em) rectangle (-0.2em,-0.2em);}
\textbf{LSTM-based.}
The paper~\cite{rw_lstm} proposes a sequence learning approach to detect weakness in the opcode of smart contracts. 
Particularly, this paper uses one-hot encoding and an embedding matrix to represent the contract's opcode. 
The obtained code vectors are used as input to train an LSTM model for determining whether the given smart contract is safe or vulnerable (i.e., binary classification). 

The LSTM-based scheme yields limited detection performance since: (i) The reported F1 score of $86\%$ is relatively low. We hypothesize that this might be because different vulnerability types have diverse behaviors, thus making it hard to distinguish the group of multiple vulnerabilities from the safe contracts. (ii) The LSTM model only provides a binary decision about contract security without distinguishing vulnerability types.

\textbf{AWD-LSTM based.} A sequence-based multi-class classification scheme is presented in~\cite{gogineni2020multi}. This paper adapts `Average Stochastic Gradient Descent Weighted Dropped LSTM' (AWD-LSTM)~\cite{merity2017regularizing} for vulnerability detection. 
The proposed model consists of two parts: a pre-trained encoder for language tasks~\cite{howard2018universal}, and an LSTM-based classifier for vulnerability classification. 
This method works on the opcode-level and can detect three vulnerability types. 

Compared to \sys{}, the AWD-LSTM based detection method has the following constraints: \textit{(i) Non-uniform effectiveness.} The multi-class detection performance of~\cite{gogineni2020multi} is not uniformly effective across different vulnerabilities. In particular,~\cite{gogineni2020multi} yields an F1 score of $95\%$ on safe contracts and $30\%$ on Prodigal contracts~\cite{nikolic2018finding}. \sys{} features a much smaller performance divergence across different classes as can be seen from Table~\ref{tab:class-metrics}. \textit{(ii) Not extensible.} The AWD-LSTM based model in~\cite{gogineni2020multi} is a fixed design to detect pre-specified vulnerability types. The extension to incorporate new attacks is not considered in~\cite{gogineni2020multi}.

\vspace{0.1em}
\textbf{CNN-based.} 
The paper~\cite{huang2018hunting} transforms the contract bytecode into fix-sized RGB color images and trains a convolution neural network for vulnerability detection. 

Similarly to \sys{}, CNN-based classifier uses multi-label classification, which has a low confidence score when determining the exact vulnerability types. %The inspection time of a single contract is $1.5$ seconds.} 

Compare to our work, the CNN-based detection scheme has the following limitations: (i) The multi-label classification performance is not satisfying due to its low confidence level. We hypothesize that this is because image representation of the bytecode and the CNN architecture ignore the sequential information existing in the contract. (ii) The extensibility/generalization ability of the CNN-based detection method is neither discussed nor evaluated. 

\textbf{GNN-based.} The paper~\cite{zhuangsmart} proposes a graph neural network (GNN)-based approach. In particular, this work builds a `contract graph' from the contract's source code where nodes and edges represent critical function calls/variables and temporal execution trace, respectively. This graph is normalized to highlight important nodes and passed to a temporal message propagation (TMP) network for vulnerability detection.

While supporting multi-class detection, the GNN-based method has the following drawbacks: \textit{(i) Restricted applicability.} The paper~\cite{zhuangsmart} requires to build a graph from the contract's source code. However, the source code is typically hard to obtain from the public blockchain. \sys{} operates on the contract bytecode which is publicly available from the EVM, thus can be deployed in more scenarios. \textit{(ii) Limited effectiveness.} The GNN-based technique yields an average F1 score of $77\%$ across all three vulnerabilities. We hypothesis that the graph normalization process in~\cite{zhuangsmart} does not preserve the malicious nodes responsible for vulnerability exploitation, leading to the low F1 score. 
\vspace{-0.7em}
\section{Conclusion}
\label{sec:conclusion}
% \vspace{-0.3em}

To ensure the safety of the Ethereum cryptocurrency system,  
we present \sys{}, the first deep learning-based automated framework that supports concurrent detection of multiple vulnerability classes and lightweight transfer learning. 
We identify two key components of vulnerability detection: feature extraction of general smart contracts and each particular vulnerability class, and disentangle these two subtasks.
\sys{}'s multi-output RNN design is highly modular, scalable, efficient, and extensible as opposed to the previous works.
Empirical results show that \sys{} achieves an average of $95\%$ detection accuracy in terms of F1 score across various vulnerability classes and can be quickly adapted on the new vulnerability data. Given an unknown smart contract, \sys{} can provide parallel detection of eight vulnerabilities in $0.02$ second.
As a separate contribution, we devise \datatool{}, a toolchain for dataset construction and labeling based on smart contracts' bytecode downloaded from Ethereum blockchain and existing vulnerability detection tools.
We will open source
\datatool{} toolchain and our dataset to promote research in this area.

\clearpage
\appendix
\section{Appendix}
\label{sec:appendix}
In this section, we provide a detailed introduction about cryptocurrency systems and illustrate how we perform dataset chunking for \sys{}'s training. 

\subsection{Ethereum Platform}
\label{sec:appendix:ethereum} 
Blockchain is proposed as a distributed ledger that records transactions between two parties in a verifiable and permanent way~\cite{zheng2017overview}. 
Ethereum is an open-sourced cryptocurrency platform based on blockchain and provides a Turing-complete Ethereum Virtual Machine (EVM) that enables developers to deploy decentralized applications. 
A cryptocurrency platform has the following key characteristics:

% \vspace{0.2em}
% {\tikz\draw[black,fill=black] (-0.5em,-0.5em) rectangle (-0.2em,-0.2em);} 
\textbf{Decentralized Nature.} 
In contrast to conventional currencies, virtual money is not administered by a central authority but by a distributed peer-to-peer network. A network of nodes, the so-called \textit{`miners'}, are responsible to perform money transactions, data storage, and updates~\cite{ethereum-whitepaper}. 
Note that on a blockchain, all code, data, and transactions are shared and available for inspection on every single node. 
All actions performed inside of this network need to be confirmed by the majority of all participating nodes~\cite{ethereum-background-01}.  

% \vspace{0.2em}
% {\tikz\draw[black,fill=black] (-0.5em,-0.5em) rectangle (-0.2em,-0.2em);}
\textbf{Mathematical Algorithm as a Basis of Cryptocurrency \\Value.} Money, in the Ethereum context called Ether, can be initially earned by solving a complex mathematical problem that can be accepted by the other nodes. The process is called \textit{mining}. 
Once a transaction is triggered, a state value for the new block is calculated and verified by the participating nodes in the network~\cite{ethereum-whitepaper}.
The amount of money available in the network is also ensured to be limited. 
As such, the Ether and the associated owners can be tracked at all times~\cite{ethereum-background-02}.

% \noindent \textbf{Resilience to Data Manipulations from Outside:} 
% \vspace{0.2em}
% {\tikz\draw[black,fill=black] (-0.5em,-0.5em) rectangle (-0.2em,-0.2em);} 
\textbf{Resilience to Data Manipulations from Outside.} 
The information of the mined money is stored in a public data structure, i.e., the blockchain. 
More specifically, modifications and transactions are stored in blocks that are appended to the chain in a chronological order after checking its validity~\cite{ethereum-background-03}.
Afterward, the network rejects any attempts to alter the blockchain entries. 
Therefore, the data is immutable and irreversible~\cite{ethereum-background-02}.

% \vspace{0.2em}
% {\tikz\draw[black,fill=black] (-0.5em,-0.5em) rectangle (-0.2em,-0.2em);}
\textbf{Pseudonymous Nature.} 
In general, registration is not required to use cryptocurrency. Users that perform transactions inside the network are identified by a public key and a private key. All transactions are associated with the addresses instead of explicit users~\cite{ethereum-whitepaper}. Therefore, it is hard to determine the identity of a user although all transactions are stored publicly on the blockchain.

\subsection{Data Chunking Process}
\label{sec:appendix:chunking}
As mentioned in the challenge (C1) of Section~\ref{sec:challenges}, bytecode files contain long sequences, thus consuming large memory during DNN training. To solve this challenge, the collected bytecode data needs to be partitioned. 
Note that the pre-labeling process discussed in Section~\ref{sec:implementation:preclassification} is based on the smart contract as a whole. It is not guaranteed that vulnerabilities will occur in the chunked segments of a smart contract since direct partition of the contract removes the inherent dependency traces of potential attacks. As such, we decide to chunk the dataset constructed in Section~\ref{sec:data} while keeping the bytecode of each contract complete to relax the memory usage of \sys{}'s DNN model. 

\begin{figure}[ht!]
  \centering
  \includegraphics[width=0.98\columnwidth]{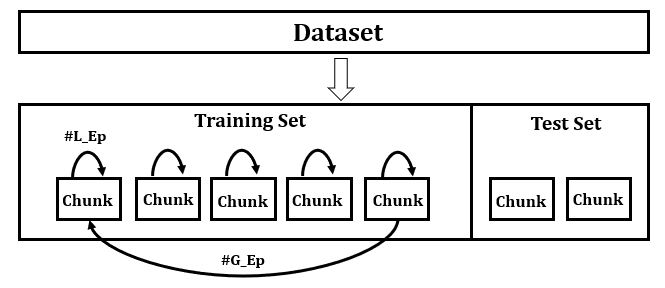}
  \vspace{-0.5em}
  \caption{Illustration of our dataset chunking method.}
  \label{fig:chunking_demo}
\end{figure}

We describe how we create data chunks from the labeled bytecode data in Section~\ref{sec:implementation:dl}. 
Figure~\ref{fig:chunking_demo} shows how DNN training is performed on the data chunks.
Two hyper-parameters are specified for model training purpose: Local\_Epoch and Global\_Epoch. As shown in the figure, \textit{Local\_Epoch} defines how often each chunk is used for training before switching to the next chunk, \textit{Global\_Epoch} defines how often the union of all the data chunks should be iterated over during the entire training process.

\vspace{-1em}

%% --------------------
%% |   Bibliography   |
%% --------------------
\bibliographystyle{ACM-Reference-Format}

\bibliography{00_main}

%% ----------------
%% |   Appendix   |
%% ----------------
%\cleardoublepage
%\input{appendix}

\end{document}